\documentclass[11pt]{article} 

\usepackage[utf8]{inputenc} 

\usepackage{geometry} 

\usepackage{graphicx} 

\usepackage[utf8]{inputenc} 
\usepackage{etex}
\usepackage{graphicx}
\usepackage{dcolumn}
\usepackage{amssymb,amsthm,amscd,amsbsy,array}
\usepackage{amstext}
\usepackage{amsmath,dsfont}
\usepackage{cancel}
\usepackage{soul} 
\usepackage{graphics,xcolor}
\usepackage{bm}
\usepackage{booktabs} 
\usepackage{array} 
\usepackage{paralist} 
\usepackage{verbatim} 
\usepackage{subfig} 
\usepackage{authblk} 
\usepackage{bbding} 

\usepackage{multirow}

\usepackage[numbers,sort&compress]{natbib}

\numberwithin{equation}{section}

\let\ssection=\section
\renewcommand{\section}{\setcounter{equation}{0}\ssection}
\textheight=
230mm 
\newcommand{\half}{{\frac{1}{2}}}
\def\2{{\half}}

\def\beq{\begin{equation}}
\def\eeq{\end{equation}}
\def\beqa{\begin{eqnarray}}
\def\eeqa{\end{eqnarray}}

\def\barray{\left(\begin{array}}
\def\earray{\end{array}\right)}
\def\barraynb{\begin{array}}
\def\earraynb{\end{array}}

\def\smallover#1/#2{\hbox{$\textstyle\frac{#1}{#2}$}} %

\usepackage[colorlinks=true, pdfstartview=FitV, linkcolor=blue, citecolor=blue, urlcolor=blue]{hyperref} 
\usepackage{color}

\newcommand{\gb}{\colorbox{green}}

\def\benu{\begin{enumerate}}
\def\eenu{\end{enumerate}}

\def\?{{\;\gb{\,?\,} \;}}

\newcommand{\os}[2]{{\overset{\,\scalebox{0.5}{($#2$)}}{#1}}{}}

\usepackage{fancyhdr} 
\usepackage{sectsty}
\allsectionsfont{\sffamily\mdseries\upshape} 

\usepackage[nottoc,notlof,notlot]{tocbibind} 
\usepackage[titles,subfigure]{tocloft} 

\title{The $1/c$ expansion of general relativity in a $3+1$ formulation, revisited}

\author[]{Mahmut Elbistan\thanks{mahmut.elbistan@bilgi.edu.tr } }
\affil[]{\vspace{0.2cm}\small Department of Energy Systems Engineering, Istanbul Bilgi University, Istanbul 34060, T\"urkiye}

\date{} 

\begin{document}

\maketitle

\begin{abstract}
We study the  $1/c$ expansion of general relativity within a formulation that is compatible with both the Arnowitt-Deser-Misner and the Kol-Smolkin decompositions. The Einstein-Hilbert action takes a common form for those decompositions as they are dual to each other. We first develop a method to expand this generic form without choosing a particular slicing and then push the expansion up to $c^{-3}$ order within this novel approach. Next, we apply our technique to the Arnowitt-Deser-Misner decomposition and expand it up to $c^{-3}$ order explicitly. In order to demonstrate the applicability of our method and to highlight the duality at the level of expansion, we also perform the expansion in the Kol-Smolkin decomposition up to $c^{-1}$ order.  Lastly, we make some all-order observations.
\vspace{0.5cm}


\end{abstract}

\pagebreak

\tableofcontents

\newpage
\section{Introduction}

Einstein's theory of general relativity (GR) unifies geometry of space-time and its matter content in an elegant way. However, the non-linear nature of its field equations make it difficult to find exact solutions within the theory \cite{Stephani:2003tm}. Therefore, various approximation schemes have been developed. Among them, there is a particularly successful one called the Post-Newtonian (PN) expansion \cite{Dautcourt:1996pm, Tichy:2011te, PoissonWill} which was built in terms of Newton-Cartan geometry. As being a weak field and slow velocity approximation, this expansion scheme is non-relativistic (NR). Apart from the effective field theory formalisms (see for instance \cite{Damour1}), there is also Post-Minkowskian approximation \cite{Damour2} which is an approximation in the gravitational constant without assuming small velocities. The $1/c$ expansion, which is the subject of this paper, is another NR approximation and it extends the PN expansion as it allows the strong field interactions \cite{VandenBleeken:2017rij, VandenBleeken:2019gqa, Hansen:2019pkl, Hansen:2020pqs, Hansen:2020wqw, Ergen:2020yop}. It is an expansion scheme in the inverse powers of the speed of light $c$. Let us note that, in addition to the $1/c$ expansion (also known as the large $c$ expansion), there is a small $c$ expansion called the Carroll expansion of GR \cite{Hansen:2021fxi, Gomis:2019nih} which  is again NR. For the interested reader, we refer to the growing literature on the subject \cite{Cariglia:2018hyr, Bergshoeff:2022eog, Concha:2022jdc, Hartong:2022lsy, Ecker:2025ncp}.  

Recently, in \cite{Elbistan:2022plu} the large $c$ expansion of GR was worked out within a $3+1$ formulation\footnote{See \cite{Gourgoulhon:2012ffd} for a review of $3+1$ formalism of GR.}. In addition to the standard Arnowitt-Deser-Misner (ADM) decomposition \cite{Arnowitt:1962hi}, the related formulation is also suitable for its dual which was dubbed by the authors the Kol-Smolkin (KS) decomposition\footnote{See the original work of Kol and Smolkin  \cite{Kol:2010si}. For an earlier use of that decomposition we refer to \cite{Deriglazov:2017jub}. }. Here, the duality means that the roles of the metric and its inverse get interchanged between those decompositions. Despite this fact, they have a common form of the Einstein-Hilbert action. Having showed that, the authors  in \cite{Elbistan:2022plu} confined themselves to the KS decomposition and performed the expansion up to  $c^{-2}$ order. They also investigated the symmetries and the equations of motion in the expansion within the KS formulation. 

In this paper, we take a step back and begin by asking the question whether the duality of the decompositions hold at the level of expansion. In order to understand this, we return to the original parent action which is valid for both decompositions and  invent a new method to perform the expansion without choosing a specific decomposition. As a result, the idea of duality leads to a novel and systematic expansion scheme that generates compact formulas up to $c^{-3}$ order which are applicable to both decompositions readily. Since the KS form and its expansion were heavily explored in \cite{Elbistan:2022plu}, we apply our generic results to the ADM case and expand the action up to $c^{-3}$ order which goes beyond the former results.  

We emphasize that our expansion scheme includes both odd and even powers of $1/c$. A systematic method of obtaining the $1/c^2$ expansion (sometimes called the even expansion)  which includes only even powers of $1/c$ was built in \cite{Elbistan:2022plu}. Combining the method therein and our results here, one may push the even expansion up to the order $c^{-6}$. In this paper, based on the idea of duality, our aim is to devise a unified method for the $1/c$ expansion without referring to a specific decomposition and extend the computations beyond those previous results. 

Let us mention some novelties and advantages of our formalism. First of all, we build our expansion framework on the  concept of duality which is an exciting and fruitful idea in physics. That enables us to devise an algorithm which is quicker and more transparent. Secondly, our formulae are valid for both decompositions. Therefore, they are applicable to the situations wherever ADM and KS decompositions are chosen. Yet, our arguments are not restricted to those cases. Any set of dual decompositions and the related $1/c$ expansion can be studied with the help of our framework. Lastly, our computations dealing with the dual decompositions might shed some light on the relation between Galilean and Carroll expansions of GR.

The outline of the paper is the following: In Section \ref{parentact}, we introduce the dual formulations and their common Einstein-Hilbert action which were found in \cite{Elbistan:2022plu}. In addition, we briefly recall the usual $1/c$ expansion scheme.  In Section \ref{intstep}, we present our new unified perspective on the expansion of the parent action of the ADM and the KS decompositions. In Section \ref{identities}, we quickly generate novel identities which help us to perform the expansion at the desired orders. In Section \ref{Ricciexpansion}, we start the expansion procedure by applying our identities.  In Section \ref{stepsexp}, we complete the expansion of the parent action up to $c^{-3}$ order. In Section \ref{ADMexpansion}, we apply our findings to the ADM decomposition and expand the related action up to the order of $c^{-3}$ explicitly. In Section \ref{furthers}, we apply our formalism to the KS decomposition and expand it up to $c^{-1}$ to highlight the role of the duality in the expansion. We also comment on the possible terms at any order for both decompositions.   In Section \ref{discuss}, we discuss our results and possible applications of the theory. In Appendices \ref{Kerrdec}, \ref{tdersteps}, \ref{Poisson} and \ref{NC} we provide some details, information about the even expansion scheme and the Newton-Cartan split of our theory.

\section{Parent action for ADM and KS formulations and the $1/c$ expansion}
\label{parentact}

\subsection{Parent action and its ingredients}

Referring to \cite{Elbistan:2022plu} for details, here we provide the basic information that we use\footnote{See our Appendix \ref{NC} for relation to the Newton-Cartan framework. }.  
Let us begin with the two decompositions that we are interested in. 
The ADM formalism contains the lapse function $N(t, \vec{x})$, the shift vector $N^i(t, \vec{x})$ and the $d$-dimensional spatial metric $h_{ij}(t, \vec{x})$. A Lorentzian metric tensor $g_{\mu\nu}$ is decomposed in terms of those elements as\footnote{In this paper, we mainly consider $d=3$.}
\begin{equation}
\label{ADMm}
ds^2 = -c^2 N^2 dt^2 + h_{ij}(dx^i + cN^i dt)(dx^j + cN^j dt),
\end{equation}
where $i, j= 1,\cdots , d$. Above $N(t, \vec{x})$ is the scalar field, $N^{i}(t, \vec{x})$ is the spatial vector and $h_{ij}(t, \vec{x})$ is the spatial metric. 
On the other hand, the KS coordinates yields a different decomposition of the same metric with other fields $M(t, \vec{x})$, $M_i(t, \vec{x})$ and $h_{ij}(t, \vec{x})$ as
\begin{equation}
\label{KSm}
ds^2=- M^2(cdt + M_i dx^i)(cdt + M_j dx^j) + h_{ij} dx^i dx^j.
\end{equation} 
The spatial metric $h_{ij}$ and its inverse $h^{ij}$ lowers and raises the indices, respectively and they satisfy   $h^{ij}h_{jk}= \delta^i_k$. Although we have used the same symbol, we shall note that  $h_{ij}$ is different for each decomposition. The relation between KS and ADM fields is
$$
N^2 -h_{ij}^{ADM} N^i N^j = M^2, \quad    h_{ij}^{ADM} N^j = -M^2 M_i ,  \quad h_{ij}^{ADM} = h_{ij}^{KS} - M^2M_i  M_j. 
$$
A typical illustration is the decomposition of the Kerr metric in our Appendix \ref{Kerrdec}.

One can easily confirm that  the inverse of the KS metric (\ref{KSm}) is in the same form with the ADM metric (\ref{ADMm}): these decompositions are dual to each other. Remarkably, they have a common form of the Einstein-Hilbert action \cite{Elbistan:2022plu} which decomposes as
\begin{equation}
\label{SEH}
S_{EH} = \int R[g] \sqrt{-g}\ d^{d+1} x= \int \Big(\hat{R} + K_{ij}K^{ij} - K^2  \Big) n_t \sqrt{h} d^{d+1} x.
\end{equation}
The Ricci scalar $R[g]$ is the combination $R[g]= \hat{R} + K_{ij}K^{ij} - K^2  $ and $\sqrt{-g} = n_t \sqrt{h}$ where $\sqrt{h}$ is the square root of the determinant of the spatial metric $h_{ij}$.
Above, $n_t$ turns out to be the scalar field $N(t, \vec{x})$ in the ADM and $M(t, \vec{x})$ in the KS formalism. 
$K(t, \vec{x})$ is the trace of the $2$-tensor $K_{ij}(t, \vec{x})$, i.e.,  $K = h^{ij}K_{ij}$ for both decompositions. They are composed of derivatives of the fields and for each decomposition we have a different expression, see (\ref{KADM}) and (\ref{KKS}).  On the other hand, $\hat{R}$ is the $d$-dimensional Ricci scalar defined in terms of hatted covariant derivatives and hatted Christoffels  
\begin{eqnarray}
\label{hatRij}
\hat{R}_{ij} &=& \hat{D}_k \hat{\Gamma} ^k_{ij} - \hat{D}_j \hat{\Gamma} ^k_{ik} - \hat{\Gamma}^k_{km} \hat{\Gamma}^m_{ij} + \hat{\Gamma}^m_{ik} \hat{\Gamma}^k_{jm}, \\ 
 \hat{R} =  h^{ij}\hat{R}_{ij} &=&  (h^{ij}h^{kl} - h^{ik}h^{jl}) h_{ln}(\hat{D}_k \hat{\Gamma} ^n_{ij} - \hat{\Gamma}^n_{km} \hat{\Gamma}^m_{ij} ).
\label{hatR0}
\end{eqnarray}
For both decompositions those Christoffel symbols are expressed in terms of the hatted derivatives $\hat{\partial}_i = e^\mu_i \partial_\mu$ with $\mu = 1, \cdots , d+1$,
\begin{equation}
\label{Christoffel}
\hat{\Gamma}^k_{ij} = \frac{1}{2} h^{km} (\hat{\partial}_i h_{mj} + \hat{\partial_j } h_{mi} - \hat{\partial}_m h_{ij}).
\end{equation}
The frame vectors $e_i = e^\mu_i \partial_\mu$ are defined as in (\ref{ADMchoice}) and (\ref{KSchoice}) for the ADM and the KS forms, respectively. See also our Appendix \ref{NC}.

Although the $d$-dimensional metric $h_{ij}$ plays different roles in (\ref{ADMm}) and (\ref{KSm}), it is covariantly constant in both decompositions
\begin{equation}
\label{covid}
\hat{D}_i h_{jk} = \hat{\partial}_i h_{jk} - \hat{\Gamma}^m_{ij} h_{mk} - \hat{\Gamma}^m_{ik}h_{im} = 0.
\end{equation} 
Depending on the decomposition, the above identity (\ref{covid}) may contain different orders of $1/c$. This is because the derivative $\hat{\partial}_i = e^\mu_i \partial_\mu$ may or may not include the time derivative accompanied by a power of $c^{-1}$. And for each decomposition the definition of  the hatted Christoffels $\hat{\Gamma}^i_{jk}$ (\ref{Christoffel}) differs with various inverse powers of $c$, see (\ref{ADMchoice}) for the ADM case and (\ref{KSChristoffel}) for the KS case.  Therefore, the identity (\ref{covid}) can be separated into subidentities including different orders for each decomposition. In the following sections, we will show that as a generic identity (\ref{covid}) is quite useful.  

The action (\ref{SEH}) has the same form, but the meanings of the individual elements $\hat{R}, K_{ij}, n_t$ and their orders are different in each decomposition. In order to show this, let us note the particular forms of the tensor $K_{ij}(t, \vec{x})$ explicitly. While it is symmetric for the ADM case
\begin{equation}
\label{KADM}
K_{ij}^{ADM} = N^{-1}(\frac{1}{2}\dot{h}_{ij} - \hat{D}_{(i} N_{j)}),  \quad   \hat{D}_{(i} N_{j)} = \frac{1}{2}\Big(  \hat{D}_i N_j + \hat{D}_j N_i \Big), 
\end{equation}
this is  no longer true for KS form 
\begin{equation}
\label{KKS}
K_{ij}^{KS}=M^{-1}\left( \frac{1}{2} \dot{h}_{ij} - M^2 \hat{\partial}_{[i}M_{j]} \right), \quad    \hat{\partial}_{[i}M_{j]}=\frac{1}{2}(\hat{\partial}_i M_j - \hat{\partial}_j M_i) ,
\end{equation}
where the dot $\dot{}$ means $\partial_t$.
Both tensors contain different $1/c$ orders because of that time derivatives.  Therefore, the expansion of these terms are worked out for each decomposition separately to keep track of that powers in the usual expansion scheme (see \cite{Elbistan:2022plu} and the references therein).

Here, we do not follow that usual approach. Starting directly from the common Einstein-Hilbert action (\ref{SEH}), our aim is to develop an expansion scheme for both decompositions at once. In other words, we will devise a new technique which handles the large $c$ expansion relying on the generic expressions like the identity (\ref{covid}). But first, let us briefly remind the elements of the usual expansion scheme.

\subsection{The usual expansion scheme}
\label{usualexp}

Now, we will quickly outline the expansion procedure in \cite{Elbistan:2022plu}.
The original parent action (\ref{SEH}) has terms with $1/c^0, 1/c, 1/c^2$ orders as it contains derivatives up to second order, see our Appendix \ref{tdersteps} for more details. Moreover,  the definitions of the terms in the action and their orders depend on the specific decomposition, as we have observed by comparing (\ref{KADM}) with (\ref{KKS}). The actual terms and their orders will be determined only if a specific decomposition is adopted.

Therefore, in the usual expansion scheme one first begins by choosing a decomposition. For instance in  \cite{Elbistan:2022plu}, we immediately chose the KS decomposition of the unexpanded Lagrangian (\ref{KSm}). Grouping the terms with the same order, we expanded the related Lagrangian (\ref{SEH}) by assuming that the fields are analytic 
\begin{equation}
\label{KSexp}
M = \os{M}{0} + \frac{1}{c} \os{M}{1}+ \cdots ,\quad
M^i = \os{M}{0}^i + \frac{1}{c} \os{M}{1}^i + \cdots , \quad
h_{ij} = \os{h}{0}_{ij} + \frac{1}{c} \os{h}{1}_{ij}+\cdots ,
\end{equation}
with the convention $\os{M}{0} = M$, $ \os{M}{0}^i  = M^i$ and $\os{h}{0}_{ij} = h_{ij}$ to simplify the notation. 
The order is indicated upon the fields. Thus, in the usual expansion scheme the inverse powers of $c$ can be  traced via summing over those indices and the number of time derivatives each accompanied with an inverse power of $c$. Such a procedure requires a great care and effort as it is performed one by one to each term while keeping track of the orders.

On the other hand, one may study within the ADM decomposition (\ref{ADMm}) of the same metric, as well. In that case, we expand the associated fields as 
\begin{equation}
\label{ADMexp}
N = \os{N}{0} + \frac{1}{c} \os{N}{1}+ \cdots ,\quad
N^i = \os{N}{0}^i + \frac{1}{c} \os{N}{1}^i + \cdots , \quad
h_{ij} = \os{h}{0}_{ij} + \frac{1}{c} \os{h}{1}_{ij}+\cdots ,
\end{equation}
where we may set $\os{N}{0} = N$, $ \os{N}{0}^i  = N^i$ and $\os{h}{0}_{ij} = h_{ij}$ again. 

ADM or KS, the idea of having a parent action (\ref{SEH}) for both decompositions is blurred by that specific choice. In this paper, we would like to put the idea of duality of the action forward and preserve it even at the level of the expansion. 
Instead of limiting ourself to a particular slicing, we will pave our own way and establish a new technique to deal with the large $c$ expansion based on the generic form of the action (\ref{SEH}). The stumbling block on our way is the fact that 
the definitions of the terms $\hat{R}$, $K_{ij}$, ... are changing for each decomposition with different powers of $1/c$. 
Even within the same decomposition, we have to keep track of the orders with care. We explain our expansion technique in the next section.

\section{A novel framework to expand the parent action step by step}
\label{intstep}

As outlined above, in the usual expansion scheme one first decides on the particular slicing i.e., ADM (\ref{ADMm}) or KS (\ref{KSm}) and then classifies the terms in the related Lagrangian (\ref{SEH}) according to their inverse powers of $c$. Having distinguished between the terms with different orders,  only in the third step the expansion procedure (\ref{KSexp}) or (\ref{ADMexp})  is initiated. The last step is to compute and to collect the terms in the desired order of the expanded Lagrangian. 

In our novel framework, we will reverse this procedure. Without specifying a particular decomposition (ADM or KS) and without distinguishing between the different orders in the unexpanded action (\ref{SEH}), we propose a new expansion scheme where we directly start from the parent action (\ref{SEH}) and perform an expansion scheme valid for both decompositions. Specific order of a specific decomposition will be chosen only at the end. 

Our technique makes use of the generic identities like (\ref{covid}) and we even develop more of them. Therefore, the duality of ADM and KS forms will be used at full power. When it comes to fields, we follow the usual expansion scheme outlined as in Section \ref{usualexp}. However, as we do not specify the decomposition at the beginning, our expanded Lagrangians will contain 3 different orders because of the derivatives\footnote{For a proof see our Appendix \ref{tdersteps}.}.  

In Section \ref{Matryoshka}, we introduce our framework which does not refer to any particular decomposition. 
Then, we generate identities including various orders within this framework in Section \ref{identities}.  In Section \ref{Ricciexpansion}, we start expanding the Ricci tensor within our formulation using the aforementioned identities.

\subsection{Matryoshka dolls}
\label{Matryoshka}

The orders of the terms in the parent Lagrangian (\ref{SEH}) 
\begin{equation}
\label{LEH}
L = R\sqrt{-g} = (\hat{R} + K_{ij}K^{ij} - K^2 ) n_t \sqrt{h}
\end{equation}
cannot be determined unless we adopt a specific slicing. For our aim, we would like to devise a method that can be directly applied to (\ref{LEH}) so that we can expand both ADM and KS decompositions within the same framework.  

Let us spell out our novel method that we call the \emph{steps}. In this framework the terms in the Lagrangian (\ref{LEH}) are expanded to the next order altogether. Therefore, each step will contain terms with three different inverse powers of $c$ rooted from the derivatives in (\ref{LEH}) and include all the terms of the preceding steps. Our step expansion is schematically established within a nested structure reminding us the matryoshka dolls. 

For instance, the \emph{zeroth step} $ [R\sqrt{-g}]_0$ refers to the unexpanded Lagrangian (\ref{LEH}). As its terms contain at most 2 derivatives, at this order we have terms with orders $1/c^0, 1/c, 1/c^2$. 
The \emph{first step} $ [R\sqrt{-g}]_1$ in the expansion amounts to shift the pieces of the zeroth step Lagrangian to the next order. Therefore, at that step we will capture terms with orders $1/c, 1/c^2, 1/c^3$. 
By iteration, the \emph{nth  step} $ [R\sqrt{-g}]_n$ contains terms with $1/c^n, 1/c^{n+1}, 1/c^{n+2}$ orders.   

Schematically our expansion scheme is the following 
\begin{equation}
[R\sqrt{-g}] \to [R\sqrt{-g}]_0 + [R\sqrt{-g}]_1 + [R\sqrt{-g}]_2 +\cdots
\end{equation}
where its nested structure is obvious as each step contains all the elements of its predecessors 
\begin{subequations}
\label{Mdolls}
\begin{align}
\label{Mdollsa}
[L]_0&= [R\sqrt{-g}]_0 = [R]_0  [\sqrt{-g}]_0, \\
\label{Mdollsb}
[L]_1& =[R\sqrt{-g}]_1 = [R]_1 [\sqrt{-g}]_0 + [R]_0 [\sqrt{-g}]_1, \\
\label{Mdollsc}
[L]_2&= [R\sqrt{-g}]_2 = [R]_2 [\sqrt{-g}]_0 + [R]_1[\sqrt{-g}]_1 + [R]_0[\sqrt{-g}]_2, \\
\label{Mdollsc3}
[L]_3&= [R\sqrt{-g}]_3 = [R]_3 [\sqrt{-g}]_0 + [R]_2[\sqrt{-g}]_1 + [R]_1[\sqrt{-g}]_2 + [R]_0[\sqrt{-g}]_3,  \\
\label{Mdollsd}
[L]_n&= [R\sqrt{-g}]_n = [R]_n [\sqrt{-g}]_0 + [R]_{n-1}[\sqrt{-g}]_1 + \cdots + [R]_1[\sqrt{-g}]_{n-1} + [R]_0[\sqrt{-g}]_n.
\end{align}
\end{subequations}   

We observe that for each step $m$ we need to compute only 2 novel terms $[R]_m$ and $[\sqrt{-g}]_m$ since the rest is contained in the previous steps.  For example, it is enough to compute $[R]_1$ and $[\sqrt{-g}]_1$ for the first step (\ref{Mdollsb}) because the elements $[\sqrt{-g}]_0$ and $[R]_0$ should be known from (\ref{Mdollsa}). Moreover, the term $[\sqrt{-g}]_m =[ n_t \sqrt{h}]_m$ is easy to deal with as it contains only fields but not derivatives. See (\ref{gmeasure}) for the first three orders of it. 

Thus, our procedure boils down to calculating $[R]_m = [\hat{R} + K_{ij}K^{ij} - K^2 ]_m$ for each step $m$. As it is seen from (\ref{KADM}) and (\ref{KKS}), $[ K_{ij}K^{ij} - K^2]_m$ part heavily depends on the decomposition. We can only make some formal computations about that part and will be able to finalize them only after we select the decomposition.  

However, the computation of the non-trivial $[\hat{R}]_m$ is quite systematic through the steps outlined in (\ref{Mdolls}) and the results are promising. As we will show in further sections, we can successfully conjecture and even compute it at the desired order (which is $c^{-3}$ here).  The gain is that our procedure is valid for both decompositions: having calculated $[\hat{R}]_m$ one can select, for instance the ADM decomposition at the end and find out the expanded Lagrangian with a little effort. The difficulty is that one seems to deal with extra terms for each order in our framework. 
More precisely, in order to obtain the action expanded at $1/c^n$ order, we need to collect the associated terms from the steps $n-2$, $n-1$ and $n$. 

And it is exactly this point where our goal of generating identities will be of use. This novel framework allows us to derive generic identities based on (\ref{covid}) for each  step $m$. These identities will be extremely helpful to compute $[\hat{R}]_m$. 

Below, we will derive those identities which also serves as a good introduction to our method.

\section{Identities for each step and the expansion }
\label{identities}

\subsection{Expansion of the inverse metric and the determinant}

The identity  $h^{ij}h_{jk} =\delta^i_k$ should be valid even after expansion  (\ref{KSexp}) and (\ref{ADMexp}). This idea helps us to generate the expansion of the inverse metric $h^{ij}$ as
\begin{equation} 
(h_{ij} + a_{ij} +b_{ij} + c_{ij}+...)(h^{jk} + A^{jk} + B^{jk} + C^{jk}+...) = \delta^i_k,
\end{equation}
where we set $\os{h}{1}_{ij} = a_{ij}, \os{h}{2}_{ij} = b_{ij}, \os{h}{3}_{ij} = c_{ij}$ for the first three orders.

The corresponding expansion terms of the inverse metric are found as
\begin{subequations}
\label{invh}
\begin{align}
\label{invh1}
A^{ij} &= - h^{ik}a_{kl}h^{lj} = -a^{ij}, \\
\label{invh2}
B^{ij} &=  - b^{ij}+ a^i_k a^{kj}, \\
\label{invh3}
C^{ij} &= - c^{ij} + a^i_k b^{kj} + b^i_ka^{kj} - a^{ik}a_{km}a^{mj}.
\end{align}
\end{subequations}
Based on the above results, we can expand the determinant under the square root, i.e.,  $\sqrt{h}$. In addition to $[\sqrt{-g}]_0 = \sqrt{h}n_t $, for the first three terms in the expansion we have\footnote{In order to avoid excessive usage of ``overset'', hereafter we remove the index for zeroth order fields.} 
\begin{subequations}
\label{gmeasure}
\begin{align}
\label{gmeasure1}
[\sqrt{-g}]_1 &=  \sqrt{h} \left(\os{n}{1}_t + \frac{1}{2} h^{ij}a_{ij}\right) , \\
\label{gmeasure2}
[\sqrt{-g}]_2 &=  \sqrt{h} \left( \os{n}{2}_t +  \frac{\os{n}{1}_t}{2} h^{ij} a_{ij} + \frac{n_t}{2} h^{ij}b_{ij}+ \frac{n_t}{8}(h^{ij} a_{ij})^2 - \frac{n_t}{4}  a^{ij}a_{ij} \right), \\
\label{gmeasure3}
[\sqrt{-g}]_3 &=  \sqrt{h} \left( \os{n}{3}_t  + \frac{\os{n}{2}_t}{2}h^{ij}a_{ij}+ \frac{\os{n}{1}_t}{2} h^{ij}b_{ij} +  \frac{\os{n}{1}_t}{8}(h^{ij} a_{ij})^2 - \frac{\os{n}{1}_t }{4} a^{ij}a_{ij} \right) \nonumber \\
&+ \sqrt{h} \ \frac{n_t}{2}\left( h^{ij}c_{ij} - a^{ij}b_{ij} + \frac{1}{3} a^{ij}a_j^k a_{ki} +\frac{1}{2} h^{ij}a_{ij} (h^{kl}b_{kl} -1/2 a^{kl} a_{kl}) + \frac{1}{24} (h^{ij}a_{ij})^3  \right) ,
\end{align}
\end{subequations}
where $\sqrt{h} \equiv \os{\sqrt{h}}{0}$. As we move further in the expansion, the terms are gettting more and more complicated. In \cite{Elbistan:2022plu}, it was shown that one can get rid of trace terms $h^{ij} \os{h}{n}_{ij} =0$ when $n\geq 1$ by a choice of coordinate using the time redefinition symmetry. Hereafter, we adopt this condition in (\ref{gmeasure}).

The expansion of the spatial metric and its determinant is the usual one. In the following, we will discuss the expansion of the terms which contain not only the fields but also their time derivatives.

\subsection{Covariant derivative of the metric}

There is a second identity (\ref{covid}) valid both for ADM and KS decompositions at the zeroth step where no expansion has been performed. Now, we show that it can be used to generate other identities for the further steps in a novel way. Although being covariant, such identities shall include expanded terms with various orders. 

Let's begin with the first step of  (\ref{covid})
\begin{equation}
\label{Dh1}
[\hat{D}_i h_{jk}]_1 = \hat{D}_i \os{h}{1}_{jk} + \os{e}{1}^\mu_{i}\partial_\mu h_{jk} - [\hat{\Gamma}^m_{ij}]_1h_{mk}- [\hat{\Gamma}^m_{ik}]_1 h_{jm} =0. 
\end{equation}
The above equation contains both the expansion of the fields like $ \os{h}{1}_{jk} = a_{jk}$ (\ref{invh1}) with a definite order and the steps of the Christoffel symbols $[\hat{\Gamma}^m_{ij}]_1$ which may contain various powers of $1/c$
depending on the decomposition. As we would like to keep our formulation free of decompositions at this level, (\ref{Dh1}) is very welcome. In fact, it is nothing but an elegant way of expressing $[\hat{\Gamma}^i_{jk}]_1$ in terms of elementary fields as
\begin{equation}
\label{defGamma1v2}
[\hat{\Gamma}^l_{ij}]_1  = \frac{1}{2} h^{lk} \big(\hat{D}_i \os{h}{1}_{jk} +   \hat{D}_j \os{h}{1}_{ik} -\hat{D}_k \os{h}{1}_{ij} +\os{e}{1}^\mu_i\partial_\mu h_{jk} + \os{e}{1}^\mu_j\partial_\mu h_{ik} - \os{e}{1}^\mu_k\partial_\mu h_{ij}     \big).
\end{equation}
Similarly, taking one more step we get
\begin{equation}
\label{Dh2}
[\hat{D}_i h_{jk}]_2 =\hat{D}_i \os{h}{2}_{jk} +  \os{e}{1}^\mu_i \partial_\mu \os{h}{1}_{jk} +  \os{e}{2}^\mu_i \partial_\mu h_{jk} - [\hat{\Gamma}^m_{ij}]_1 \os{h}{1}_{mk} -  [\hat{\Gamma}^m_{ik}]_1 \os{h}{1}_{jm} -  [\hat{\Gamma}^m_{ij}]_2 h_{mk} - [\hat{\Gamma}^m_{ik}]_2 h_{jm} =0,
\end{equation}
which allows us to make the definition of $[\hat{\Gamma}^l_{ij}]_2 $ as
\begin{eqnarray}
\label{defGamma2v2}
[\hat{\Gamma}^l_{ij}]_2  = \frac{1}{2} h^{lk} &\Big(&\hat{D}_i \os{h}{2}_{jk} +   \hat{D}_j \os{h}{2}_{ik} -\hat{D}_k \os{h}{2}_{ij}  +\os{e}{2}^\mu_i\partial_\mu h_{jk} + \os{e}{2}^\mu_j\partial_\mu h_{ik} - \os{e}{2}^\mu_k\partial_\mu h_{ij}  \nonumber \\
 &+&\os{e}{1}^\mu_i\partial_\mu\os{h}{1}_{jk} + \os{e}{1}^\mu_j\partial_\mu \os{h}{1}_{ik} - \os{e}{1}^\mu_k\partial_\mu \os{h}{1}_{ij} - 2 \os{h}{1}_{mk} [\hat{\Gamma}^m_{ij}]_1\Big), 
\end{eqnarray}
where $[\hat{\Gamma}^l_{ij}]_1$ has already been given explicitly in (\ref{defGamma1v2}).
By iteration, we can even make an all-order observation such that 
\begin{equation}
\label{Dhn}
[\hat{D}_i h_{jk}]_n = \hat{D}_i \os{h}{n}_{jk} +  \os{e}{1}^\mu_i \partial_\mu\os{h}{n-1}_{jk}+\cdots +   \os{e}{n}^\mu_i \partial_\mu h_{jk} -[\hat{\Gamma}^m_{ij}]_1 \os{h}{n-1}_{mk} -\cdots - [\hat{\Gamma}^m_{ij}]_n h_{mk} -[\hat{\Gamma}^m_{ik}]_1 \os{h}{n-1}_{mj} -\cdots - [\hat{\Gamma}^m_{ik}]_n h_{mj} =0,
\end{equation}
which, for any $n \geq 1$,  can be inverted to define $ [\hat{\Gamma}^l_{ij}]_n $  as
\begin{equation}
\label{defGamman}
 [\hat{\Gamma}^l_{ij}]_n = h^{kl}\Big( \hat{D}_i \os{h}{n}_{jk} + \os{e}{1}^\mu_i\partial_\mu \os{h}{n-1}_{jk}+\cdots +  \os{e}{n-1}^\mu_i\partial_\mu h_{jk}  -[\hat{\Gamma}^m_{ij}]_1 \os{h}{n-1}_{mk} -\cdots - [\hat{\Gamma}^m_{ij}]_{n-1} \os{h}{1}_{mk} -[\hat{\Gamma}^m_{ik}]_1 \os{h}{n-1}_{mj} -\cdots - [\hat{\Gamma}^m_{ik}]_n h_{mj} \Big)
\end{equation}
At first glance the definition (\ref{defGamman}) seems weird because the right hand side contains $ [\hat{\Gamma}^m_{ik}]_n$.
Like (\ref{defGamma1v2}) and (\ref{defGamma2v2}), one can use it to define $[\hat{\Gamma}^l_{ij}]_n$ in terms of elementary fields. 
 However,  when it comes to its derivation
\begin{eqnarray}
\label{DGamman}
\hat{D}_k [\hat{\Gamma}^p_{ij}]_n = h^{sp} \hat{D}_k&\Big(&\hat{D}_i \os{h}{n}_{js} + \os{e}{1}^\mu_i\partial_\mu \os{h}{n-1}_{js}+\cdots +  \os{e}{n}^\mu_i\partial_\mu h_{js}\nonumber \\  &-&[\hat{\Gamma}^m_{ij}]_1 \os{h}{n-1}_{ms} -\cdots - [\hat{\Gamma}^m_{ij}]_{n-1} \os{h}{1}_{ms} -[\hat{\Gamma}^m_{is}]_1 \os{h}{n-1}_{mj} -\cdots - [\hat{\Gamma}^m_{is}]_n h_{mj} \Big), \quad \quad \quad 
\end{eqnarray}
the definition (\ref{defGamman}) becomes very useful.  This is because during the expansion (\ref{Mdolls}) it will be contracted with the Wheeler-DeWitt metric $h^{ij} h^{kl} - h^{ik}h^{jl}$. As a result that weird part is eliminated due to symmetry arguments, leaving us with simpler expressions.

Let us note the definitions that we will use in the further sections,
\begin{subequations}
\begin{align}
\label{hatDgamma1}
\hat{D}_k [\hat{\Gamma}^l_{ij}]_1 &= h^{sl} \Big(  \hat{D}_k (\hat{D}_i \os{h}{1}_{js}) - h_{jm} \hat{D}_k  [\hat{\Gamma}^m_{is}]_1 + \hat{D}_k ( \os{e}{1}^\mu_i\partial_\mu h_{js}) \Big), \\
\label{hatDgamma2}
\hat{D}_k [\hat{\Gamma}^l_{ij}]_2 &= h^{sl} \Big(  \hat{D}_k (\hat{D}_i \os{h}{2}_{js}) - h_{jm} \hat{D}_k [\hat{\Gamma}^m_{is}]_2 + \hat{D}_k (\os{e}{1}^\mu_i\partial_\mu \os{h}{1}_{js} + \os{e}{2}^\mu_i\partial_\mu h_{js} ) - \hat{D}_k \big( [\hat{\Gamma}^m_{ij}]_1  \os{h}{1}_{ms} +  [\hat{\Gamma}^m_{is}]_1 \os{h}{1}_{jm} \big) \Big),  \qquad \\
\label{hatDgamma3}
\hat{D}_k [\hat{\Gamma}^l_{ij}]_3 &= h^{sl} \hat{D}_k \Big( \hat{D}_i \os{h}{3}_{js} - h_{jm} [\hat{\Gamma}^m_{is}]_3 +\os{e}{1}^\mu\partial_\mu\os{h}{2}_{js} +\os{e}{2}^\mu\partial_\mu\os{h}{1}_{js} +\os{e}{3}^\mu\partial_\mu h_{js} \nonumber \\
 &  \qquad  \qquad - [\hat{\Gamma}^m_{ij}]_1\os{h}{2}_{ms} - [\hat{\Gamma}^m_{ij}]_2\os{h}{1}_{ms} -  [\hat{\Gamma}^m_{is}]_1\os{h}{2}_{mj} - [\hat{\Gamma}^m_{is}]_2\os{h}{1}_{mj}\Big),
\end{align}
\end{subequations}
where we have already used $\hat{D}_k h_{ij} = \hat{D}_k h^{ij}=0$.

\section{Ricci tensor and its expansion}
\label{Ricciexpansion}

\subsection{nth step of the Ricci tensor}

As seen from our nested structure (\ref{Mdolls}) an arbitrary step $n$ of the expanded Lagrangian
\begin{eqnarray} 
[L]_n = [R\sqrt{-g}]_n = [(\hat{R} + K_{ij} K^{ij} - K^2)\sqrt{-g}]_n
\end{eqnarray}
contains only two genuine elements $[R]_n$ and $[\sqrt{-g}]_n$ to be calculated. The latter can be expanded via the usual expansion scheme (\ref{gmeasure}).  The difficult part is the nth step of the Ricci scalar $[R]_n$ 
\begin{equation}
\label{Riccin}
[R]_n \equiv [\hat{R} + K_{ij} K^{ij} - K^2]_n = [\hat{R}]_n + [K_{ij} K^{ij} - K^2]_n,
\end{equation}
where our technique presented in the following section will be essential. Putting $[K_{ij} K^{ij} - K^2]_n$ aside for now, let us consider the first term  $[\hat{R}]_n$
\begin{eqnarray}
\label{hatRn}
[\hat{R}]_n = [h^{ij}\hat{R}_{ij}]_n = \os{h}{n}^{ij}[\hat{R}_{ij}]_0 +   \os{h}{n-1}^{ij}[\hat{R}_{ij}]_1 +\cdots +  h^{ij}[\hat{R}_{ij}]_n,
\end{eqnarray}
which is common for both decompositions.  
To compute it  first we need $[\hat{R}_{ij}]_n$ for any $n \geq 1$ and  then contract it with the inverse metrics in the expansion (\ref{invh}).  We achieved the former via expanding (\ref{hatRij}),
\begin{eqnarray}
\label{hRijn}
[\hat{R}_{ij}]_n &=& \hat{D}_k [\hat{\Gamma}^k_{ij}]_n - \hat{D}_j [\hat{\Gamma}^k_{ik}]_n \nonumber \\
&+& \os{e}{n}^\mu_k\partial_\mu \hat{\Gamma}^k_{ij} + \cdots +\os{e}{1}^\mu_k\partial_\mu [\hat{\Gamma}^k_{ij}]_{n-1} -  \os{e}{n}^\mu_j\partial_\mu \hat{\Gamma}^k_{ik} + \cdots +\os{e}{1}^\mu_j\partial_\mu [\hat{\Gamma}^k_{ik}]_{n-1} \nonumber \\
&+& [\hat{\Gamma}^k_{km}]_{n-1}  [\hat{\Gamma}^m_{ij}]_{1} +\cdots + [\hat{\Gamma}^k_{km}]_{1}  [\hat{\Gamma}^m_{ij}]_{n-1}  - [\hat{\Gamma}^m_{ik}]_{n-1}  [\hat{\Gamma}^k_{jm}]_{1} -\cdots - [\hat{\Gamma}^m_{ik}]_{1}  [\hat{\Gamma}^k_{jm}]_{n-1}.  \qquad \quad
\end{eqnarray} 
Akin to (\ref{defGamman}) and (\ref{DGamman}), $[\hat{R}_{ij}]_n $ (\ref{hRijn}) is again an all-order expression showing the power of our method. Right now it is generic and valid for both decompositions.  
Let us spell out first 3 steps of (\ref{hRijn}) 
\begin{subequations}
\label{hatRij123}
\begin{align}
\label{hatRij1}
[\hat{R}_{ij}]_1 &= \hat{D}_k [\hat{\Gamma}^k_{ij}]_1 - \hat{D}_j [\hat{\Gamma}^k_{ik}]_1  +\os{e}{1}^\mu_k\partial_\mu \hat{\Gamma}^k_{ij} - \os{e}{1}^\mu_j \partial_\mu \hat{\Gamma}^k_{ik}, \\
\label{hatRij2}
[\hat{R}_{ij}]_2 &= \hat{D}_k [\hat{\Gamma}^k_{ij}]_2 - \hat{D}_j [\hat{\Gamma}^k_{ik}]_2  +\os{e}{2}^\mu_k\partial_\mu \hat{\Gamma}^k_{ij} - \os{e}{2}^\mu_j \partial_\mu \hat{\Gamma}^k_{ik} +\os{e}{1}^\mu_k\partial_\mu [\hat{\Gamma}^k_{ij}]_1 - \os{e}{1}^\mu_j \partial_\mu [\hat{\Gamma}^k_{ik}]_1 \nonumber\\ &+ [\hat{\Gamma}^k_{km}]_1[\hat{\Gamma}^m_{ij}]_1 - [\hat{\Gamma}^m_{ik}]_1[\hat{\Gamma}^k_{jm}]_1, \\
\label{hatRij3}
 [\hat{R}_{ij}]_3 &= \hat{D}_k [\hat{\Gamma}^k_{ij}]_3 - \hat{D}_j [\hat{\Gamma}^k_{ik}]_3 +\os{e}{3}^\mu_k\partial_\mu \hat{\Gamma}^k_{ij} - \os{e}{3}^\mu_j \partial_\mu \hat{\Gamma}^k_{ik}    +\os{e}{2}^\mu_k\partial_\mu [\hat{\Gamma}^k_{ij}]_1 - \os{e}{2}^\mu_j \partial_\mu [\hat{\Gamma}^k_{ik}]_1 +\os{e}{1}^\mu_k\partial_\mu [\hat{\Gamma}^k_{ij}]_2 - \os{e}{1}^\mu_j \partial_\mu [\hat{\Gamma}^k_{ik}]_2 \nonumber \\
&+ [\hat{\Gamma}^k_{km}]_2[\hat{\Gamma}^m_{ij}]_1 - [\hat{\Gamma}^m_{ik}]_2[\hat{\Gamma}^k_{jm}]_1 +[\hat{\Gamma}^k_{km}]_1[\hat{\Gamma}^m_{ij}]_2 - [\hat{\Gamma}^m_{ik}]_1[\hat{\Gamma}^k_{jm}]_2, 
\end{align}
\end{subequations}
which we will use in the next sections.

\subsection{Identities and the first 3 steps}

The real difficulty about (\ref{hatRij123}) is the computation of the covariant derivatives $ \hat{D}_k [\hat{\Gamma}^k_{ij}]_n$. But, they will be eventually contracted with $(h^{ij}h^{kl} - h^{ik}h^{jl}) $ which considerably simplifies the calculations. 

For instance, when $n=1$  we need to contract $[\hat{R}_{ij}]_1$ with $h^{ij}$ (\ref{hatRn}) which turns out to be
\begin{eqnarray}  
\label{hhatR1}
h^{ij}[\hat{R}_{ij}]_1 &=& (h^{ij} h^{kl} - h^{ik} h^{jl}) h_{ln}\Big( \hat{D}_k [\hat{\Gamma}^n_{ij}]_1 + \os{e}{1}^\mu_k\partial_\mu \hat{\Gamma}^n_{ij}  \Big), \nonumber \\
&=&  (h^{ij} h^{kl} - h^{ik} h^{jl}) \Big( \hat{D}_k \hat{D}_i \os{h}{1}_{jl} + \hat{D}_k (\os{e}{1}^\mu_i \partial_\mu h_{jl}) +h_{ln} \os{e}{1}^\mu_k\partial_\mu\hat{\Gamma}^n_{ij} \Big), 
\end{eqnarray}
where we used (\ref{hatRij1}) in the first line and  (\ref{hatDgamma1}) in the second line. Pay attention to the fact that the problematic part with $[\hat{\Gamma}^m_{il}]_1$ is eliminated when contracted with the Wheeler-DeWitt metric due its  $i\leftrightarrow l$ antisymmetry. 

The $n=2$ case involves $h^{ij} [\hat{R}_{ij}]_2$ which contains $1/c^2, 1/c^3 $ and $1/c^4$ orders,
\begin{eqnarray}
\label{hhatR2}
h^{ij}[\hat{R}_{ij}]_2 &=& (h^{ij} h^{kl} - h^{ik} h^{jl}) h_{ln}\Big(\hat{D}_k [\hat{\Gamma}^n_{ij}]_2 + \os{e}{2}^\mu_k\partial_\mu \hat{\Gamma}^n_{ij}     + \os{e}{1}^\mu_k\partial_\mu [\hat{\Gamma}^n_{ij}]_1 +  [\hat{\Gamma}^n_{km}]_1 [\hat{\Gamma}^m_{ij}]_1   \Big), \nonumber \\
&=&  (h^{ij} h^{kl} - h^{ik} h^{jl})  \left(   \hat{D}_k \hat{D}_i \os{h}{2}_{jl} + \hat{D}_k \big(\os{e}{2}^\mu_i \partial_\mu h_{jl} + \os{e}{1}^\mu_i \partial_\mu \os{h}{1}_{jl} \big)  - \hat{D}_k \big( \os{h}{1}_{lm}[\hat{\Gamma}^m_{ij}]_1 \big)  \right) \nonumber \\
&+& (h^{ij} h^{kl} - h^{ik} h^{jl}) h_{ln}\left(   \os{e}{2}^\mu_k\partial_\mu \hat{\Gamma}^n_{ij} + \os{e}{1}^\mu_k\partial_\mu [\hat{\Gamma}^n_{ij}]_1 +[\hat{\Gamma}^n_{km}]_1 [\hat{\Gamma}^m_{ij}]_1 \right),
\end{eqnarray}
where we used the definition (\ref{hatRij2}) in the first line and (\ref{hatDgamma2}) and the symmetry properties of the Wheeler-DeWitt metric in the further lines. Let us note that $n=2$ case  (\ref{hatRn}) requires also the contraction 
\begin{equation}
- a^{ij} [\hat{R}_{ij}]_1 = - (a^{ij} h^{kl} - a^{ik}h^{jl}) h_{ln}(\hat{D}_k [\hat{\Gamma}^n_{ij}]_1 + \os{e}{1}^\mu_k\partial_\mu \hat{\Gamma}^n_{ij}).
\end{equation}  
When combined with (\ref{hhatR2}), its broken  $i\leftrightarrow l$ antisymmetry is restored and our identity (\ref{hatDgamma1}) still simplifies the calculations in a beautiful manner.

In order to go beyond $c^{-2}$ order,  we need to calculate  $h^{ij} [\hat{R}_{ij}]_3$ by contracting Wheeler-DeWitt metric and (\ref{hatRij3}) which results in\footnote{Pay attention to the fact that (\ref{hhatR3}) contains terms with orders $c^{-3}, c^{-4}, c^{-5}$.}
\begin{eqnarray}
\label{hhatR3}
h^{ij}[\hat{R}_{ij}]_3 &=& (h^{ij} h^{kl} - h^{ik} h^{jl})  \left(  \hat{D}_k \hat{D}_i \os{h}{3}_{jl} +\hat{D}_k \big(  \os{e}{3}^\mu_i\partial_\mu h_{jl} + \os{e}{2}^\mu_i\partial_\mu \os{h}{1}_{jl}+ \os{e}{1}^\mu_i\partial_\mu \os{h}{2}_{jl} \big)  - \hat{D}_k \big( \os{h}{2}_{lm} [\hat{\Gamma}^m_{ij}]_1 + \os{h}{1}_{lm} [\hat{\Gamma}^m_{ij}]_2   \big)    \right) \nonumber \\
&+& (h^{ij} h^{kl} - h^{ik} h^{jl}) h_{ln}\left(    \os{e}{3}^\mu_k\partial_\mu \hat{\Gamma}^n_{ij} + \os{e}{2}^\mu_k\partial_\mu [\hat{\Gamma}^n_{ij}]_1+ \os{e}{1}^\mu_k\partial_\mu [\hat{\Gamma}^n_{ij}]_2+ [\hat{\Gamma}^n_{km}]_2 [\hat{\Gamma}^m_{ij}]_1 + [\hat{\Gamma}^n_{km}]_1 [\hat{\Gamma}^m_{ij}]_2  \right).
\end{eqnarray}

Note that most of the computations have already been done in (\ref{hhatR1}) ,  (\ref{hhatR2}) and  (\ref{hhatR3}). And we achieved those results without confining ourselves to a specific decomposition: the above results can be applied to both the ADM and the KS forms and we do that in our Sections \ref{ADMexpansion} and \ref{furthers}. 

There are a few remaining computations, though.  For instance, in (\ref{hhatR3}) one still needs to deal with the terms like $\hat{D}_k \big( \os{h}{2}_{lm} [\hat{\Gamma}^m_{ij}]_1 + \os{h}{1}_{lm} [\hat{\Gamma}^m_{ij}]_2 \big)$. At the end of the next section, when we deal with $[\hat{R}]_3$ which is nothing but (\ref{hatRn}) with $n=3$, the expression (\ref{hhatR3}) will be combined with $(-b^{ij} + a^{ik}a_k^j) [\hat{R}_{ij}]_1$ and $-a^{ij} [\hat{R}_{ij}]_2$. This combination restores the  $i\leftrightarrow l$ antisymmetry to cancel the related expressions.

\section{Steps and the expansion of the parent Lagrangian}
\label{stepsexp}

Equipped with the necessary formulas above, now we pass to the expansion stage in our framework.

\subsection{Step 0}

The zeroth step Lagrangian (\ref{Mdollsa}) is the smallest doll and it is identical to the unexpanded one (\ref{LEH}) with (\ref{hatR0})
\begin{equation}
\label{step0L}
[L]_0 =  [R\sqrt{-g}]_0 = [R]_0 [\sqrt{-g}]_0=  n_t \sqrt{h} (h^{ij} h^{kl} - h^{ik} h^{jl})  \left(  h_{ln}(\hat{D}_k \hat{\Gamma}^n_{ij} - \hat{\Gamma}^n_{km}\hat{\Gamma}^m_{ij}) - K_{ij}K_{kl}  \right), 
\end{equation}
where $ [R]_0$ is the original Ricci scalar and $[\sqrt{-g}]_0 =  n_t \sqrt{h}$. This term contains 3 lowest orders which are $c^0, c^{-1}, c^{-2}$.

\subsection{Step 1}

As noted in (\ref{Mdollsb}),  the first step Lagrangian $[L]_1$ contains $ [R]_0$ and $[\sqrt{-g}]_0 $ of its predecessor (\ref{step0L}). $[\sqrt{-g}]_1$ is readily noted in (\ref{gmeasure1}). Substituting $n=1$  in (\ref{Riccin}) and subsequently in (\ref{hatRn}), we obtain
\begin{eqnarray}
\label{step1L}
[L]_1 &=&  [R\sqrt{-g}]_1 = [R]_1 [\sqrt{-g}]_0 +  [R]_0 [\sqrt{-g}]_1,  \\
&=& \sqrt{h} \left( n_t \Big( - a^{ij} \hat{R}_{ij} + h^{ij} [\hat{R}_{ij}]_1  - \big[(h^{ij} h^{kl} - h^{ik} h^{jl})K_{ij}K_{kl}\big]_1 \Big)  + \os{n}{1}_t \Big( h^{ij} \hat{R}_{ij} -(h^{ij} h^{kl} - h^{ik} h^{jl})K_{ij}K_{kl} \Big) \right) \nonumber .
\end{eqnarray}
We still need to compute  $[\hat{R}]_1 = [h^{ij} \hat{R}_{ij} ]_1 = \os{h}{1}^{ij} \hat{R}_{ij} + h^{ij}[\hat{R}_{ij}]_1$ where $ \os{h}{1}^{ij} = - a^{ij}$. 
 Note that it is different for each decomposition and it contains different inverse powers of $c$.  Remarkably, such generic forms valid for both decompositions can be handled within our novel framework
\begin{eqnarray}
\label{hatR1}
[\hat{R}]_1 &=& [h^{ij}\hat{R}_{ij}]_1 = \os{h}{1}^{ij}\hat{R}_{ij} + h^{ij}[\hat{R}_{ij}]_1, \\
&=&  - (a^{ij}h^{kl} - a^{ik}h^{jl} ) h_{ln} (\hat{D}_k \hat{\Gamma}^n_{ij} - \hat{\Gamma}^n_{km}\hat{\Gamma}^m_{ij}) +  (h^{ij}h^{kl} - h^{ik}h^{jl}) \left(\hat{D}_k \hat{D}_i a_{jl} + \hat{D}_k (\os{e}{1}^\mu_i\partial_\mu h_{jl}) + h_{ln} \os{e}{1}^\mu_k\partial_\mu \hat{\Gamma}^n_{ij} \right)\nonumber,
\end{eqnarray}
where we have used (\ref{DGamman}) and the $i\leftrightarrow l$ antisymmetry of  the Wheeler-DeWitt metric.   

Adding the contributions of the term $(h^{ij} h^{kl} - h^{ik} h^{jl}) K_{ij} K_{kl}$ and $ \big[(h^{ij} h^{kl} - h^{ik} h^{jl})K_{ij}K_{kl}\big]_1$ we finally get our $[L]_1$ as
\begin{eqnarray}
\label{step1Lfin}
[L]_1 &=&  \sqrt{h}  n_t   (h^{ij} h^{kl} - h^{ik} h^{jl})  \big( \hat{D}_k \hat{D}_i a_{jl} + \hat{D}_k (\os{e}{1}^\mu_i\partial_\mu h_{jl}) + h_{ln} \os{e}{1}^\mu_k\partial_\mu \hat{\Gamma}^n_{ij}  -2 [K_{ij}]_1 K_{kl}  \big)   \nonumber \\
&+&  \sqrt{h}    \os{n}{1}_t  (h^{ij} h^{kl} - h^{ik} h^{jl})\big( h_{ln} (  \hat{D}_k \hat{\Gamma}^n_{ij}  - \hat{\Gamma}^n_{km} \hat{\Gamma}^m_{ij}) - K_{ij} K_{kl}   \big) \nonumber \\
&-&  \sqrt{h}  n_t  (a^{ij} h^{kl} - a^{ik} h^{jl}) \Big(   h_{ln}( \hat{D}_k \hat{\Gamma}^n_{ij} - \hat{\Gamma}^n_{km} \hat{\Gamma}^m_{ij})  - K_{ij} K_{kl} + K_{lj} K_{ki} \Big). \quad 
\end{eqnarray}
Step 1 Lagrangian (\ref{step1Lfin}) is really effective. Its first line contains the second covariant derivative of the expanded field $\os{h}{1}_{ij} = a_{ij}$ and derivatives of $h_{ij}$ itself. Derivative of the Christoffels (\ref{Christoffel}) can easily be found. Second line is nothing but the step 0 Lagrangian (\ref{step0L}) multiplied with the expanded field $ \os{n}{1}_t$ of the measure and in the third line we have just expanded the Wheeler-DeWitt metric\footnote{There might be additional cancellations due to our coordinate choice $h^{ij} a_{ij}=0$ but to keep our argument general we did not remove them.}. 

The expanded Lagrangian (\ref{step1Lfin}) reflects the duality at the level of expansion. Thus, one may apply it to  both (\ref{ADMm}) and (\ref{KSm}) by substituting the related fields easily because almost all the terms are readily computed before selecting a specific decomposition. Therefore, it is easy to keep track of the order of $1/c$. Different decompositions have different forms of $K_{ij}$, therefore we leave the explicit computation to next chapters.  See the first order ADM Lagrangian in Section \ref{ADM1} and its counterpart KS Lagrangian in Section \ref{KS1}.

\subsection{Step 2}
\label{step2detail}

On the other hand, the second step Lagrangian $[L]_2$ (\ref{Mdollsc}) requires the second step Ricci scalar $[R]_2$ and the metric determinant $ [\sqrt{-g}]_2 $. Apart from these, the other elements were already contained in the zeroth (\ref{step0L}) and the first (\ref{step1L}) steps. As $ [\sqrt{-g}]_2 $ is noted in (\ref{gmeasure2}), we now focus on $[R]_2 = [\hat{R} + K_{ij}K^{ij} - K^2]_2$. 

Putting $n=2$ in (\ref{hatRn}), we get
\begin{equation}
[\hat{R}]_2 = B^{ij} \hat{R}_{ij} + A^{ij}[\hat{R}_{ij}]_1 + h^{ij}[\hat{R}_{ij}]_2,
\end{equation}
where the inverse expanded metrics $A^{ij}$ and $B^{ij}$ are given in (\ref{invh1}) and (\ref{invh2}), respectively. The last term $h^{ij}[\hat{R}_{ij}]_2$ (\ref{hhatR2}) can be further improved as
\begin{eqnarray}
h^{ij}[\hat{R}_{ij}]_2 &=& (h^{ij} h^{kl} - h^{ik} h^{jl}) \big(  \hat{D}_k \hat{D}_i b_{jl} + \hat{D}_k ( \os{e}{2}^\mu_i\partial_\mu h_{jl} + \os{e}{1}^\mu_i\partial_\mu a_{jl}) + h_{ln} \os{e}{2}^\mu_k\partial_\mu\hat{\Gamma}^n_{ij} +\os{e}{1}^\mu_k\partial_\mu(h_{lm} [\hat{\Gamma}^m_{ij}]_1)   \big) \nonumber\\ 
&-&  (h^{ij} h^{kl} - h^{ik} h^{jl}) \big(  [\hat{\Gamma}^m_{ij}]_1 [\hat{\Gamma}^n_{kl}]_1 h_{mn} - a_{lm} \hat{D}_k[\hat{\Gamma}^m_{ij}]_1 \big)
\end{eqnarray}
Adding $- a^{ij} [\hat{R}_{ij}]_1$ and $B^{ij} \hat{R}_{ij}$ to it and after some calculations we obtain the desired result as
\begin{eqnarray}
 [\hat{R}]_2 &=& (h^{ij} h^{kl} - h^{ik} h^{jl}) \big(  \hat{D}_k\hat{D}_i b_{jl} + \hat{D}_k (\os{e}{2}^\mu_i \partial_\mu h_{jl} +\hat{e}{1}^\mu_i\partial_\mu a_{jl} ) + h_{lm}(\os{e}{2}^\mu_k\partial_\mu\hat{\Gamma}^m_{ij} +\os{e}{1}^\mu_k\partial_\mu[\hat{\Gamma}^m_{ij}]_1 )   \big) \nonumber \\
&+&  (h^{ij} h^{kl} - h^{ik} h^{jl}) \frac{1}{2} (\hat{D}_m a_{kl} -2 \hat{D}_k a_{lm} +\os{e}{1}^\mu_k\partial_\mu h_{kl}) [\hat{\Gamma}^m_{ij}]_1  \nonumber \\
&-& (a^{ij} h^{kl} - a^{ik}h^{jl}) \big(  \hat{D}_k \hat{D}_i a_{jl} -   \hat{D}_k \hat{D}_l a_{ij} + \hat{D}_k (\os{e}{1}^\mu_i\partial_\mu h_{jl} - \os{e}{1}^\mu_l\partial_\mu h_{ij})  +h_{lm} \os{e}{1}^\mu_k\partial_\mu \hat{\Gamma}^m_{ij} \big)\nonumber\\
&+& \big( (-b^{ij} + a^i_s a^{sj}) h^{kl} - (-b^{ik} + a^{i}_sa^{sk}) h^{jl} \big) h_{ln} (\hat{D}_k\hat{\Gamma}^n_{ij} -\hat{\Gamma}^n_{km}\hat{\Gamma}^m_{ij}). \quad
\end{eqnarray}
We complete the above expression with 
\begin{eqnarray}
[K_{ij} K^{ij} - K^2 ]_2 &=& - (h^{ij}h^{kl}- h^{ik}h^{jl}) \big(  2 [K_{ij}]_2 K_{kl} + [K_{ij}]_1 [K_{kl}]_1 \big) \nonumber \\
&+&(a^{ij} h^{kl} - a^{ik}h^{jl}) \big( [K_{ij}]_1 K_{kl} + K_{ij}[K_{kl}]_1 -[K_{kl}]_1 K_{ki} - K_{lj}[K_{ki}]_1    \big)\nonumber \\
&-& \big( (-b^{ij}+ a^{im}a_m^j)h^{kl} - (-b^{jk} + a^{jm}a_m^k)h^{jl} \big) (K_{ij} K_{kl} - K_{lj}K_{ki}) - (a^{ij}a^{kl} - a^{ik} a^{jl})K_{ij}K_{kl}, \quad  \quad \quad
\end{eqnarray}
to determine $[R]_2$ entirely. There remains to apply our findings and this will be done in Section \ref{ADM2}.

\subsection{Step 3}
\label{step3details}
We have come to the last step. $[L]_3$ (\ref{Mdollsd}) consists of several elements but many of them have already been computed in the previous steps. As $[\sqrt{-g}]_3$ is noted in (\ref{gmeasure3}), we only need to find out $[R]_3$
\begin{equation}
\label{step3R}
[R]_3 = [\hat{R}]_3 +[K_{ij}K^{ij} - K^2]_3.
\end{equation}  
First part
\begin{equation}
 [\hat{R}]_3 = h^{ij}[\hat{R}_{ij}]_3 - a^{ij}[\hat{R}_{ij}]_2 + B^{ij}[\hat{R}_{ij}]_1 + C^{ij}[\hat{R}_{ij}]_0,
\end{equation}
where $C^{ij} $ is given in (\ref{invh3}), after long and tedious calculations can be found as
\begin{eqnarray}
\label{step3hatR}
 [\hat{R}]_3 &=& (h^{ij}h^{kl} - h^{ik}h^{jl}) \Big( \hat{D}_k \hat{D}_i c_{jl} + \hat{D}_k (\os{e}{3}^\mu_i\partial_\mu h_{jl} +\os{e}{2}^\mu_i\partial_\mu a_{jl} +\os{e}{1}^\mu_i\partial_\mu b_{jl})   \Big)   \nonumber \\
&=&  (h^{ij}h^{kl} - h^{ik}h^{jl})h_{ln} \big( \os{e}{3}^\mu_i\partial_\mu \hat{\Gamma}^n_{ij} +\os{e}{2}^\mu_i\partial_\mu [\hat{\Gamma}^n_{ij}]_1 +\os{e}{1}^\mu_i\partial_\mu [\hat{\Gamma}^n_{ij}]_2    \big) \nonumber\\
&-& (h^{ij}h^{kl} - h^{ik}h^{jl})  \Big( 2 [\hat{\Gamma}^n_{kl}]_2[\hat{\Gamma}^m_{ij}]_1 h_{nm} +[\hat{\Gamma}^n_{kl}]_1[\hat{\Gamma}^m_{ij}]_1 a_{nm}  - (\os{e}{1}^\mu_k\partial_\mu a_{lm}  + \os{e}{2}^\mu_k\partial_\mu h_{lm}) [\hat{\Gamma}^m_{ij}]_1  - \os{e}{1}^\mu_k\partial_\mu [\hat{\Gamma}^m_{ij}]_2 \Big) \nonumber \\
&-& (a^{ij} h^{kl} - a^{ik} h^{jl}) \Big(  \hat{D}_k (\hat{D}_i b_{jl} + \os{e}{1}^\mu \partial_\mu a_{jl} + \os{e}{2}^\mu_i \partial_\mu h_{jl}    - i \leftrightarrow l ) + h_{ln} (\os{e}{2}^\mu_k \partial_\mu \hat{\Gamma}^n_{ij} + \os{e}{1}^\mu_k \partial_\mu [\hat{\Gamma}^n_{ij}]_1)  \Big)  \nonumber\\
&+& (a^{ij} h^{kl} - a^{ik} h^{jl}) ([\hat{\Gamma}^m_{lj}]_1 \os{e}{1}^\mu_k\partial_\mu h_{im}  - [\hat{\Gamma}^m_{ij}]_1 \os{e}{1}^\mu_k\partial_\mu h_{lm}+ 2 h_{mn} [\hat{\Gamma}^m_{ij}]_1 [\hat{\Gamma}^n_{kl}]_1 )  \nonumber\\
&-& (b^{ij}h^{kl} - b^{ik}h^{jl}) \big( \hat{D}_k (\hat{D}_i a_{jl} + \os{e}{1}^\mu_i \partial_\mu h_{jl} - i \leftrightarrow l)    + h_{ln} \os{e}{1}^\mu_k\partial_\mu \hat{\Gamma}^n_{ij}    \Big)  \nonumber\\
&+& \Big(  \frac{1}{2} (a^{ij} a^{kl} - a^{ik} a^{jl}) + (h^{ij} a^{mk}a_m^l - h^{ik}a^{mj}a_m^l )  \Big)  \hat{D}_k (\hat{D}_i a_{jl} +\os{e}{1}^\mu_i\partial_\mu h_{jl} - i\leftrightarrow l) \nonumber \\
&+&( a^{im}a_m^j h^{kl} - a^{im}a_m^kh^{jl}) h_{ln} \os{e}{1}^\mu_k\partial_\mu \hat{\Gamma}^n_{ij} + (C^{ij} h^{kl} - C^{ik} h^{jl}) h_{ln}(\hat{D}_k \hat{\Gamma}^n_{ij} - \hat{\Gamma}^n_{km}\hat{\Gamma}^m_{ij}). 
\end{eqnarray}
In a compact form the second piece becomes
\begin{eqnarray}
[K_{ij}K^{ij} - K^2]_3 &=& - 2(h^{ij}h^{kl} - h^{ik}h^{jl}) ([K_{ij}]_3 K_{kl} + [K_{ij}]_2 [K_{kl}]_1) \nonumber\\
&+& (a^{ij} h^{kl} - a^{ik}h^{jl}) \big( 2 [K_{ij}]_2 K_{kl} + [K_{ij}]_1 [K_{kl}]_1 - i \leftrightarrow l    \big) \nonumber\\
&+& \Big(   (-b^{ij}+ a^i_m a^{mj})h^{kl} - (-b^{ik} + a^i_m a^{mk})h^{jl} + \frac{1}{2} (a^{ij} a^{kl} - a^{ik} a^{jl})  \Big) ( 2 [K_{ij}]_1 K_{kl} - i\leftrightarrow l   ) \nonumber\\
&+& (C^{ij}h^{kl} - C^{ik}h^{jl} - B^{ij}a^{kl} + B^{ik}a^{jl}) (K_{ij} K_{kl} - i\leftrightarrow l).
\end{eqnarray}

We remind that both $[\hat{R}]_3$ and $[K_{ij}K^{ij} - K^2]_3$ contain terms of order $c^{-3}, c^{-4}, c^{-5}$. In the following section, we consider $3$rd order ADM Lagrangian. Both $[\hat{R}]_3$ and $[K_{ij}K^{ij} - K^2]_3$ source this Lagrangian with their terms having no time derivative. And, the previous steps will contribute it with their terms involving derivatives.

\section{Expansion of the ADM decomposition}
\label{ADMexpansion}

In this section, we discuss the large $c$ expansion of the ADM decomposition (\ref{ADMm}) within our novel framework. Recall that in \cite{Elbistan:2022plu}, it was the KS formulation (\ref{KSm}) in which the expansion was performed. We will also push the expansion up to $c^{-3}$ order which goes beyond the results of that paper. 

The choice of the ADM decomposition (\ref{ADMm}) yields
\begin{equation}
\label{ADMchoice}
e^\mu_i = \delta^\mu_i \implies \hat{\partial}_i =\partial_i, \quad \hat{\Gamma}^i_{jk} = \Gamma^i_{jk}= \frac{1}{2}h^{il}(\partial_j h_{lk} + \partial_k h_{lj} - \partial_l h_{jk} ), 
\end{equation}
thus the hat $\hat{}$ on the covariant derivative can be removed, $ \hat{D}_i = D_i$. The tensor $K_{ij}$ is totally symmetric (\ref{KADM}). Pay attention to the fact that $\Gamma^i_{jk}$ does not contain any time derivatives and  $\os{e}{n}^\mu_i =0$ for any $n\geq 1$. Expansion of Christoffel symbols $\os{\Gamma}{m}^i_{jk}$ will be defined wherever necessary. 

In the following, we will adopt above definitions and present the expanded ADM Lagrangian order by order, up to $c^{-3}$ based on our previous findings.

\subsection{ADM Lagrangian: zeroth order}
\label{ADM0}
According to our scheme (\ref{Mdollsa}), the Lagrangian (\ref{step0L}) contains all zeroth order i.e., $c^0$ terms. Adopting (\ref{ADMchoice}), we only need to extract the related terms to get the related Lagrangian which is
\begin{equation}
\label{L0ADM}
\os{L}{0}_{ADM} = N \sqrt{h} (h^{ij} h^{kl} - h^{ik} h^{jl}) \left(h_{ln} (D_k \Gamma^n_{ij} - \Gamma^n_{km} \Gamma^m_{ij}) - N^{-2} D_{(i}N_{j)} D_{(k}N_{l)}  \right).
\end{equation} 
Pay attention to the fact that (\ref{L0ADM}) does not include any time derivatives or expanded fields. In that manner, the higher order terms will be time dependent perturbations on the top of this stationary theory.

\subsection{ADM Lagrangian: first order}
\label{ADM1}

First order ADM Lagrangian $\os{L}{1}_{ADM}$ can be obtained through a two-step process. First, we should collect all that $c^{-1}$ order terms in (\ref{step1Lfin}) adopting (\ref{ADMchoice}). Then, we should remember that $\os{L}{1}_{ADM}$  receives contributions from the terms with a single time derivative of the zeroth step Lagrangian (\ref{step0L}).

Altogether we note $\os{L}{1}_{ADM}$ as
\begin{eqnarray}
\label{L1ADM}
\os{L}{1}_{ADM} &=& \sqrt{h} N (h^{ij}h^{kl} - h^{ik} h^{jl}) \left( D_k D_i a_{jl} +2N^{-2} (N^{-1} \os{N}{1} D_{(i} N_{j)} - D_{(i} \os{N}{1}_{j)} + \os{\Gamma}{1}^m_{ij}N_m ) D_{(k} N_{l)}      \right) \nonumber \\
&-&   \sqrt{h} N (a^{ij} h^{kl} - a^{ik} h^{jl}) \left( h_{ln} (D_k \Gamma^n_{ij} - \Gamma^n_{km} \Gamma^m_{ij}) -2N^{-2} D_{(i} N_{j)} D_{(k} N_{l)}   \right) \nonumber \\
&+& \sqrt{h} \os{N}{1} (h^{ij}h^{kl} - h^{ik} h^{jl})   \left( h_{ln} (D_k \Gamma^n_{ij} - \Gamma^n_{km} \Gamma^m_{ij}) -N^{-2} D_{(i} N_{j)} D_{(k} N_{l)}   \right)\nonumber \\
&+& \sqrt{h} N^{-1}  (h^{ij}h^{kl} - h^{ik} h^{jl}) \dot{h}_{ij}  D_{(k} N_{l)},
\end{eqnarray} 
where the expanded Christoffel symbol (\ref{defGamma1v2}) is 
\begin{equation}
\os{\Gamma}{1}^l_{ij} = \frac{1}{2} h^{lk} (D_i a_{jk} + D_j a_{ik} - D_k a_{ij}).
\end{equation}
The last term in (\ref{L1ADM}) is the contribution from the step zero Lagrangian (\ref{step0L}). The rest originates from our first step (\ref{Mdollsb}) which is (\ref{step1Lfin}) and it only contains first order expanded fields. 

Let us make two more comments. Firstly, it is possible to make further simplifications in (\ref{L1ADM}) remembering our gauge choice $h^{ij} a_{ij} =0$. We would like to keep it in its present general form which is suitable for other gauge choices. Secondly, in order to show both the power of our formalism and highlight the the role of duality within the expansion scheme, we also expand the KS Lagrangian up to $c^{-1}$, cf. (\ref{L1KS}).

\subsection{ADM Lagrangian: second order}
\label{ADM2}

As we mentioned, second order ADM Lagrangian $\os{L}{2}_{ADM}$ is sourced by three different steps. Main contribution comes from the $[L]_2$ which has been worked out in Section \ref{step2detail}. The formula (\ref{Mdollsc}) shows that $[L]_2$ is also composed of three terms. All these contributions do not have any time derivatives but the expanded fields.  Secondly, the terms of $[L]_1$  (\ref{step1Lfin}) with a single time derivative contribute it. Lastly, the second order terms in (\ref{step0L}) should be added.

We write all these contributions separately. Let us start with the contributions from $[L]_2$ which contains only expanded fields. We divide them into three based on the expansion of the measure   
\begin{subequations}
\label{L20}
\begin{align}
\os{L}{2}^{0} &= \os{L}{2}^{0a} + \os{L}{2}^{0b}+ \os{L}{2}^{0c}, \\
\label{L20a}
\os{L}{2}^{0a} &= \sqrt{h} N (h^{ij}h^{kl}- h^{ik} h^{jl}) \Big(  D_k D_i b_{jl} + \frac{1}{2} (D_m a_{kl} - 2 D_k a_{lm})\os{\Gamma}{1}^m_{ij} -2N^{-3} (-\os{N}{2} + N^{-1}\os{N}{1}^2)D_{(i}N_{j)}D_{(k}N_{l)}   \Big)\nonumber \\
&+ \sqrt{h} N (h^{ij}h^{kl}- h^{ik} h^{jl}) \Big(N^{-1}\os{N}{1}\big(D_{(i}\os{N}{1}_{j)} - \os{\Gamma}{1}^m_{ij}N_m\big) -  (D_{(i}\os{N}{2}_{j)} - \os{\Gamma}{1}^m_{ij}\os{N}{1}_m -\os{\Gamma}{2}^{m}_{ij}N_m  )  \Big) 2N^{-2} D_{(k}N_{l)} \nonumber\\
&-  \sqrt{h} N^{-1} (h^{ij}h^{kl}- h^{ik} h^{jl})  \Big( N^{-2}\os{N}{1}^2 D_{(i}N_{j)}D_{(k}N_{l)}      +(D_{(i}\os{N}{1}_{j)} - \os{\Gamma}{1}^m_{ij}N_m  - 2N^{-1}\os{N}{1} D_{(i}N_{j)})(D_{(k}\os{N}{1}_{l)} -\os{\Gamma}{1}^n_{kl}N_n )     \Big)   \nonumber \qquad \\
&-  \sqrt{h}N(a^{ij}h^{kl} - a^{ik}h^{jl}) \Big(D_kD_i a_{jl} - D_k D_l a_{ij} - \big(N^{-2} (N^{-1} \os{N}{1} D_{(i}N_{j)} - D_{(i}\os{N}{1}_{j)} +\os{\Gamma}{1}^m_{ij}N_m)D_{(k}N_{l)} - i\leftrightarrow l \big)  \Big)\nonumber\\
&+   \sqrt{h}N(a^{ij}h^{kl} - a^{ik}h^{jl})  \big(N^{-2} (N^{-1} \os{N}{1} D_{(k}N_{l)} - D_{(k}\os{N}{1}_{l)} +\os{\Gamma}{1}^m_{kl}N_m)D_{(i}N_{j)}- i\leftrightarrow l \big)   \nonumber \\
 &+\sqrt{h}N \big( (-b^{ij}+ a^i_ma^{mj})h^{kl} - (b^{ik}+a^i_m a^{mj})h^{jl} \big)   ( h_{ln} (D_k \Gamma^n_{ij} - \Gamma^n_{km} \Gamma^m_{ij}) + N^{-2}D_{(i} N_{j)} D_{(k} N_{l)} - i\leftrightarrow l ) \nonumber\\
&+ \sqrt{h} N^{-1}  (a^{ij}a^{kl} -a^{ik}a^{jl})  D_{(i}N_{j)}D_{(k}N_{l)}.\\
\label{L20b}
\os{L}{2}^{0b}&=  \sqrt{h} \os{N}{1} (h^{ij}h^{kl} - h^{ik} h^{jl}) \left( D_k D_i a_{jl} +2N^{-2} (N^{-1} \os{N}{1} D_{(i} N_{j)} - D_{(i} \os{N}{1}_{j)} + \os{\Gamma}{1}^m_{ij}N_m ) D_{(k} N_{l)}      \right) \nonumber \\
&-   \sqrt{h} \os{N}{1} (a^{ij} h^{kl} - a^{ik} h^{jl}) \left( h_{ln} (D_k \Gamma^n_{ij} - \Gamma^n_{km} \Gamma^m_{ij}) -2N^{-2} D_{(i} N_{j)} D_{(k} N_{l)}   \right)  \\
\label{gh}
\os{L}{2}^{0c}&=   \sqrt{h}(\os{N}{2} - \frac{1}{4} a_{ij}a^{ij}) (h^{ij} h^{kl} - h^{ik} h^{jl}) \left(h_{ln} (D_k \Gamma^n_{ij} - \Gamma^n_{km} \Gamma^m_{ij}) - N^{-2} D_{(i}N_{j)} D_{(k}N_{l)}  \right).
\end{align}
\end{subequations} 

The second part comes from the $1/c^2$ order terms of (\ref{step1Lfin}), i.e., 
\begin{eqnarray}
\os{L}{2}^{1} &=& \sqrt{h} \os{N}{1} (h^{ij}h^{kl} - h^{ik} h^{jl})N^{-2} \dot{h}_{ij} D_{(k}N_{l)} -2N^{-1}\sqrt{h}(a^{ij}h^{kl} - a^{ik}h^{jl}) \dot{h}_{ij}  D_{(k}N_{l)} .\nonumber\\
&-&  \sqrt{h}  (h^{ij}h^{kl} - h^{ik} h^{jl}) N^{-1} \Big((2N^{-1} \os{N}{1}  D_{(i}N_{j)} -  D_{(i}\os{N}{1}_{j)}+ \os{\Gamma}{1}^m_{ij}N_m)\dot{h}_{kl} - \dot{a}_{ij}  D_{(k}N_{l)} \Big) .
\end{eqnarray}  

Finally, we note the terms with two time derivatives in (\ref{L0ADM}) as
\begin{equation}
\os{L}{2}^{2} =- \sqrt{h}  (h^{ij}h^{kl} - h^{ik} h^{jl})   \frac{N^{-1}}{4} \dot{h}_{ij} \dot{h}_{kl}.
\end{equation}

The second order ADM Lagrangian is the sum of all those contributions, 
\begin{equation}
\os{L}{2}_{ADM} = \os{L}{2}^{0} +\os{L}{2}^{1} + \os{L}{2}^{2}. 
\end{equation}
Above expressions can be simplified further. However, we would like to keep them in that block form where it is really easy to determine the origin of the terms. In addition, this form is adaptable to any gauge choice. 

Finally, let us note that this second order expansion in ADM coordinates is completely new. In our previous work \cite{Elbistan:2022plu}, we expanded the KS form of the metric up to this order.

\subsection{ADM Lagrangian: third order}

The third order ADM Lagrangian $\os{L}{3}_{ADM}$ is a combination of three different terms, as its predecessors. First piece $\os{\hat{R}}{3}$ includes only expanded fields but no time derivatives and we obtain it from $[R]_3$ (\ref{step3R}). It also receives contributions  from previous steps (\ref{Mdollsd}) such that
\begin{eqnarray}
L_3 &=& \sqrt{h} \Big( \os{R}{3} N + \os{N}{1}(\sqrt{h} N)^{-1}\os{L}{2}^{0a}  +  (\os{N}{2} - 1/4 N a^{ij} a_{ij}) (\sqrt{h}\os{N}{2})^{-1}\os{L}{1}^{0b} \Big) \nonumber \\
&+& \sqrt{h}  \Big( \os{N}{3} - \frac{1}{4} \os{N}{1} a^{ij}a_{ij} + \frac{N}{2} ( - a^{ij}b_{ij} + \frac{1}{3} a^{ij}a_j^k a_{ki})     \Big) \os{L}{0}_{ADM}  . 
\end{eqnarray}
Those terms can be recovered from (\ref{L20}) as $[R]_2 = (\sqrt{h} N)^{-1}\os{L}{2}^{0a}$ (\ref{L20a}),  $[R]_1 = (\sqrt{h}\os{N}{2})^{-1}\os{L}{1}^{0b}$  (\ref{L20b}) etc. The unknown part is the $[R]_3 = [\hat{R}]_3 + [K^{ij}K_{ij}-K^2]_3$. Let us denote $\os{\hat{R}}{3}$ as the third order terms in $[R]_3$ without time derivatives so that it contains only expanded fields. Thanks to our framework, we find it quickly in a compact manner as 
\begin{eqnarray}
\os{\hat{R}}{3} &=& \sqrt{h} N (h^{ij} h^{kl} - h^{ik}h^{jl}) \Big( D_k D_i c_{jl} +\os{\Gamma}{1}^m_{ij}(\os{\Gamma}{1}^n_{kl} a_{mn} - D_k b_{ml} - D_l b_{mk} + D_m b_{kl})   \Big) \nonumber\\
&-&  \sqrt{h} N (a^{ij}h^{kl} - a^{ik} h^{jl}) \Big(  D_k(D_i b_{jl} - D_l b_{ij})  - 2 h_{mn}\os{\Gamma}{1}^m_{ij}\os{\Gamma}{1}^n_{kl}  \Big) \nonumber\\
&-&  \sqrt{h} N \Big(  (b^{ij}h^{kl} - b^{ik} h^{jl}) - \frac{a^{ij} a^{kl} - a^{ik} a^{jl}}{2} - h^{ij}a^{km}a_m^l + h^{ik}a^{jm}a_m^l \Big) D_k (D_i a_{jl} - D_l a_{ij}) \nonumber \\
&+&   \sqrt{h} N(C^{ij} h^{kl} - C^{ik} h^{jl}) h_{ln}(D_k \Gamma^n_{ij} - \Gamma^n_{km} \Gamma^m_{ij}).
\end{eqnarray}
Above formula originates from our step 3 computation (\ref{step3hatR}) where we applied (\ref{ADMchoice}). Derivatives $ D_k b_{ml}$ come from the expanded  Christoffel symbols $\os{\Gamma}{2}^i_{jk}$.
All those terms are at the order of $c^{-3}$. To complete the first part, we need the $c^{-3}$ contributions from $ [K^{ij}K_{ij}-K^2]_3$. We define $\os{K}{3}$ as the $c^{-3}$ contributions from $ [K^{ij}K_{ij}-K^2]_3$ with expanded fields only
\begin{eqnarray}
\os{K}{3} &=& \sqrt{h} (C^{ik}h^{jl} - C^{ij}h^{kl} -B^{ik}a^{jl} + B^{ij} a^{kl} ) N^{-1} (   D_{(i}N_{j)} D_{(k} N_{l)} - i\leftrightarrow l ) \nonumber \\
&-&2 \sqrt{h} N \Big( B^{ij}h^{kl} - B^{ik}h^{jl}  + \frac{1}{2} (a^{ij} a^{kl} - a^{ik} a^{jl})  \Big)  \left( \Big( N^{-2} \os{N}{1} D_{(i}N_{j)} - N^{-1} (D_{(i}\os{N}{1}_{j)} - \os{\Gamma}{1}^m_{ij}N_m)  \Big)D_{(k}N_{l)}- i\leftrightarrow l \right) \nonumber \\
&+&  \sqrt{h}  (a^{ij} h^{kl} - h^{ik} a^{jl}) \left(  2 \big(    (-N^{-2} \os{N}{2} +N^{-3} \os{N}{1}^2) D_{(i}N_{j)} -N^{-2} \os{N}{1} (D_{(i} \os{N}{1}_{j)} - \os{\Gamma}{1}^m_{ij}N_m)  \big)N^{-1} D_{(k}N_{l)} - i \leftrightarrow l         \right) \nonumber\\
&+&  \sqrt{h}  (a^{ij} h^{kl} - h^{ik} a^{jl}) 2N^{-2}\Big(  ( D_{(i} \os{N}{2}_{j)} - \os{\Gamma}{1}^m_{ij} \os{N}{1}_m -  \os{\Gamma}{2}^m_{ij} N_m   ) D_{(k}N_{l)}   - i \leftrightarrow l \Big) \nonumber\\
&+& \sqrt{h}  (a^{ij} h^{kl} - h^{ik} a^{jl}) \Big( N^{-2}\os{N}{1} D_{(i}N_{j)} \big( N^{-2}\os{N}{1}  D_{(k}N_{l)} -N^{-1}(D_{(k} \os{N}{1}_l) - \os{\Gamma}{1}^m_{kl} N_m \big)            - i \leftrightarrow l     \Big)\nonumber\\
&-&  \sqrt{h}  (a^{ij} h^{kl} - h^{ik} a^{jl}) \Big(     N^{-1}(D_{(i} \os{N}{1}_{j)} - \os{\Gamma}{1}^m_{ij}N_m) \big( N^{-2}\os{N}{1}  D_{(k}N_{l)} -N^{-1}D_{(k} \os{N}{1}_{l)} - \os{\Gamma}{1}^m_{kl} N_m \big)      - i \leftrightarrow l     \Big)\nonumber\\
&-&  2\sqrt{h}  (h^{ij} h^{kl} - h^{ik} h^{jl}) D_{(i} N_{j)}  \Big( (-N^{-2}\os{N}{3} - N^{-4}\os{N}{1}^3 + 2N^{-3} \os{N}{1} \os{N}{2}) \Big) D_{(k} N_{l)}\nonumber\\
&-&  2\sqrt{h}  (h^{ij} h^{kl} - h^{ik} h^{jl})  D_{(i} N_{j)}  \Big( (-N^{-2} \os{N}{2} + N^{-3} \os{N}{1}^2) (D_{(k}\os{N}{1}_{l)} - \os{\Gamma}{1}^m_{kl} N_m)    \Big) \nonumber \\
&-&  2\sqrt{h}  (h^{ij} h^{kl} - h^{ik} h^{jl}) D_{(i} N_{j)}  \Big(  (-N^{-2}\os{N}{1}) (D_{(k}\os{N}{2}_{l)} - \os{\Gamma}{1}^m_{kl} \os{N}{1}_m - \os{\Gamma}{2}^m_{kl} N_m)  \Big) \nonumber\\
&-&  2\sqrt{h}  (h^{ij} h^{kl} - h^{ik} h^{jl}) D_{(i} N_{j)}  \Big(N^{-1} (D_{(k}\os{N}{3}_{l)} - \os{\Gamma}{1}^m_{kl} \os{N}{2}_m - \os{\Gamma}{2}^m_{kl} \os{N}{1}_m - \os{\Gamma}{3}^m_{kl} N_m)   \Big) \nonumber \\
&-&  2\sqrt{h} N (h^{ij} h^{kl} - h^{ik} h^{jl}) D_{(i} N_{j)}  (-N^{-2}\os{N}{2} +N^{-3}\os{N}{1}^2) \Big( N^{-2} \os{N}{1}D_{(k} N_{l)} - N^{-1} (D_{(k}\os{N}{1}_{l)} - \os{\Gamma}{1}^m_{kl} N_m)   \Big) \nonumber\\
&-&  2\sqrt{h} N^{-1} (h^{ij} h^{kl} - h^{ik} h^{jl}) (D_{(k}\os{N}{1}_{l)} - \os{\Gamma}{1}^m_{kl} N_m)\Big( N^{-2} \os{N}{1}D_{(k} N_{l)} - N^{-1} (D_{(k}\os{N}{1}_{l)} - \os{\Gamma}{1}^m_{kl} N_m)   \Big) \nonumber\\ 
&-&  2\sqrt{h}  (h^{ij} h^{kl} - h^{ik} h^{jl})  (D_{(i}\os{N}{2}_{j)} - \os{\Gamma}{1}^m_{ij} \os{N}{1}_m - \os{\Gamma}{2}^m_{ij} N_m)\Big( N^{-2} \os{N}{1}D_{(k} N_{l)} - N^{-1} (D_{(k}\os{N}{1}_{l)} - \os{\Gamma}{1}^m_{kl} N_m)   \Big). 
\end{eqnarray}

The second part includes all the terms with one time derivative and for that we revisit our step 2 Lagrangian in Section \ref{step2detail}. The last part is the one with two time derivatives and we search for such terms in (\ref{step1Lfin}). 

Let us denote the piece with two derivatives as $\os{L}{3}^2$ and write it below,
\begin{eqnarray}
\os{L}{3}^2 &=& \sqrt{h}  (h^{ij} h^{kl} - h^{ik} h^{jl}) \Big( -\frac{\os{N}{1} N^2}{2} \dot{h}_{ij} +  N^{-1} (N^{-1} \os{N}{1}\dot{h}_{ij} - \dot{a}_{ij})     \Big)\frac{\dot{h}_{kl}}{2}, \nonumber\\
&+& \sqrt{h}  (a^{ij} h^{kl} - a^{ik}h^{jl}) \frac{N^{-1}}{2} \dot{h}_{ij}\dot{h}_{kl}.
\end{eqnarray}

In addition, we note the terms with one time derivative originating from (\ref{Mdollsc}) as $\os{L}{3}^1$,
\begin{eqnarray}
\os{L}{3}^1&=& \sqrt{h} (h^{ij} h^{kl} - h^{ik} h^{jl}) (\os{N}{2} - \frac{1}{4} a_{ij}a^{ij}) N^{-2} \dot{h}_{ij}D_{(k} N_{l)} +\sqrt{h} (a^{ij} h^{kl} - a^{ik} h^{jl}) \os{N}{1} (-2N^{-2} \dot{h}_{ij} D_{(k} N_{l)})\nonumber\\
&-&\sqrt{h} (h^{ij} h^{kl} - h^{ik} h^{jl}) \os{N}{1} N^{-2}\Big( ( 2 N^{-1}\os{N}{1} D_{(i} N_{j)} - D_{(i} \os{N}{1}_{j)} + \os{\Gamma}{1}^m_{ij}N_m)\dot{h}_{kl} - \dot{a}_{ij}D_{(k} N_{l)}  \Big) \nonumber\\
&+& 2\sqrt{h} (h^{ij} h^{kl} - h^{ik} h^{jl})   \Big(  N^{-3} (-\os{N}{2} + N^{-1} \os{N}{1}^2)\dot{h}_{ij}D_{(k} N_{l)} - \frac{N^{-3}\os{N}{1}}{2} ( D_{(i} \os{N}{1}_{j)}\dot{h}_{kl} + \dot{a}_{ij}D_{(k} N_{l)} - \os{\Gamma}{1}^m_{ij}\dot{h}_{kl}N_m )  \Big)\nonumber \\
&+& 2\sqrt{h} (h^{ij} h^{kl} - h^{ik} h^{jl}) \Big(  \frac{N^{-2}}{2} (\dot{b}_{ij}D_{(k} N_{l)} + D_{(i} \os{N}{2}_{j)}\dot{h}_{kl} - \os{\Gamma}{1}^m_{ij}\os{N}{1}_m\dot{h}_{kl} - \os{\Gamma}{2}^m_{ij}N_m \dot{h}_{kl} )  - \frac{N^{-4}}{2} \os{N}{1}^2 \dot{h}_{ij} D_{(k} N_{l)} \Big) \nonumber \\ 
&+& \sqrt{h} (h^{ij} h^{kl} - h^{ik} h^{jl}) \Big( N^{-2}\dot{a}_{ij} (- D_{(k} \os{N}{1}_{l)} + \os{\Gamma}{1}^m_{kl}N_m)  - N^{-3}\os{N}{1} \big(\dot{h}_{ij} (- D_{(k} \os{N}{1}_{l)} + \os{\Gamma}{1}^m_{kl}N_m)  \big)  - D_{(i} N_{j)}\dot{a}_{kl} \Big) \nonumber\\
&+& 2 N^{-1}\sqrt{h}  \big(  (- b^{ij}+a^i_m a^{mj})h^{kl} - (-b^{ik}+ a^{im}a_m^k)h^{jl}     - \frac{a^{ij}a^{kl}- a^{ik}a^{jl}}{2}  \big)\dot{h}_{ij}D_{(k} N_{l)} 
\end{eqnarray} 

To sum up, the $c^{-3}$ order expanded ADM Lagrangian is
\begin{equation}
\os{L}{3}_{ADM} = L_3 + \os{L}{3}^1 +\os{L}{3}^2,
\end{equation}
which is a novel $c^{-3}$ result.

\section{Further results}
\label{furthers}

In order to demonstrate the full power of our framework, here we present two more results: Firstly, we provide the $c^{-1}$ order KS Lagrangian. Secondly, we comment on the possible all-order terms in the large $c$ expansion.

\subsection{KS Lagrangian: first order}
\label{KS1}

Expansion in the KS decomposition was heavily discussed in \cite{Elbistan:2022plu}. Since ADM and KS decompositions have a common form of the Einstein-Hilbert action (\ref{SEH}), our framework e.g., formulas like (\ref{Mdollsb}) and (\ref{step1Lfin}) can be applied to both of them. We only need to adopt the related definitions.   
For KS decomposition, we have
\begin{equation}
\label{KSchoice}
e^\mu_i = \delta^\mu_i - C_i\delta^\mu_0,   
\end{equation}
which leads to the following separation of the Christoffel symbol as
$$
\hat{\Gamma}^i_{jk} = \Gamma^i_{jk}+ \Delta^i_{jk},
$$
where 
\begin{subequations}
\label{KSChristoffel}
\begin{align}
\label{KSGamma}
 \Gamma^i_{jk}&= \frac{1}{2}h^{il}(\partial_j h_{lk} + \partial_k h_{lj} - \partial_l h_{jk} ), \\
\Delta^i_{jk} &= -\frac{1}{2} h^{il} (C_j \dot{h}_{lk} + C_k \dot{h}_{lj} - C_l \dot{h}_{jk}).
\label{KSDelta}
\end{align}
\end{subequations}
While the first piece (\ref{KSGamma}) is exactly at the same form of its  ADM counterpart (\ref{ADMchoice}), the second part contains time derivative of the fields. Therefore, (\ref{KSDelta}) is in $c^{-1}$ order. 

Since the first order Lagrangian already contains the zeroth order one\footnote{ That is why we call them as Matryoshka dolls.} we simply note the first order expanded KS Lagrangian here as
 \begin{eqnarray}
\label{L1KS}
\os{L}{1}_{KS}&=& M\sqrt{h} (h^{ij}h^{kl} - h^{ik} h^{jl}) \Big( D_k D_i a_{jl} - 2 M \big(\os{M}{1}\partial_{[i} C_{j]} + \partial_{[i} \os{C}{1}_{j]}\big) \partial_{[k} C_{l]}    \Big) \nonumber\\
&-& M\sqrt{h} (a^{ij}h^{kl} - a^{ik} h^{jl}) \Big( h_{ln}(D_k \Gamma^n_{ij} - \Gamma^n_{km} \Gamma^m_{ij}) - M^2 ( \partial_{[i}C_{j]}\partial_{[k} C_{l]} - i\leftrightarrow l  )  \Big) \nonumber\\
&+& \os{M}{1} (h^{ij}h^{kl} - h^{ik} h^{jl}) \Big( h_{ln}(D_k \Gamma^n_{ij} - \Gamma^n_{km} \Gamma^m_{ij}) - M^2 ( \partial_{[i}C_{j]}\partial_{[k} C_{l]}\Big) \nonumber \\
&+& M\sqrt{h} (h^{ij}h^{kl} - h^{ik} h^{jl}) \Big(  h_{ln}(D_k \Delta^n_{ij} - \Gamma^n_{km} \Delta^m_{ij} + \mathcal{D}_k\Gamma^n_{ij} - \Delta^n_{km} \Gamma^m_{ij} )  + (\dot{h}_{ij} + 2M^2 C_{[i} \dot{C}_{j]}) \partial_{[k} C_{l]}      \Big), \quad  \qquad 
\end{eqnarray}
where 
$$ 
\mathcal{D}_k\Gamma^n_{ij} = - C_k \partial_t \Gamma^n_{ij} + \Delta^n_{kl}\Gamma^l_{ij} -  \Delta^l_{ki}\Gamma^n_{lj} -  \Delta^l_{kj}\Gamma^n_{il}.
$$ 
Using $\partial_t h^{ij} = -h^{ik}\dot{h}_{kl}h^{lj}$, $\dot{\Gamma}^n_{ij}$ is calculated to be
$$
\dot{\Gamma}^n_{ij} = \frac{1}{2} h^{nm} (D_i\dot{h}_{mj} + D_j \dot{h}_{mi} - D_m \dot{h}_{ij}),
$$
where 
$$
D_i\dot{h}_{mj} = \partial_i \dot{h}_{mj} - \Gamma^l_{im} \dot{h}_{lj} - \Gamma^l_{ij}\dot{h}_{lm}.
$$ 

Our result (\ref{L1KS}) is in line with our previous one in \cite{Elbistan:2022plu}.
Comparing with its dual (\ref{L1ADM}), we see that duality preserves the terms in the expansion of $[\hat{R}]_1$. As it is indicated in (\ref{step1Lfin}), differences occur because of the $K_{ij}$ tensor which is particular to the decomposition. As in (\ref{L1ADM}), the first three terms comes from the first step Lagrangian (\ref{step1Lfin}) while the last term is the $c^{-1}$ contribution originating from the unexpanded KS Lagrangian.

\subsection{An all-order observation}

In \cite{Elbistan:2022plu}, it was shown that there is a universal piece linear in $n$th order fields  in the $n$th order KS Lagrangian. This fact can be observed in a more formal way including ADM Lagrangian within our framework. Contracting $[\hat{R}_{ij}]_n$ (\ref{hRijn}), we get 
\begin{eqnarray} 
\label{norderpiece}
h^{ij} [\hat{R}_{ij}]_n &=& (h^{ij} h^{kl} - h^{ik}h^{jl}) \hat{D}_k\Big( \hat{D}_i \os{h}{n}_{jl} + \os{e}{1}^\mu_i\partial_\mu \os{h}{n-1}_{jl} +\cdots +  \os{e}{n}^\mu_i\partial_\mu h_{jl}  - [\hat{\Gamma}^m_{ij}]_1 \os{h}{n-1}_{ml} - \cdots - [\hat{\Gamma}^m_{ij}]_{n-1} \os{h}{1}_{ml}    \Big) \nonumber\\
&+& (h^{ij} h^{kl} - h^{ik}h^{jl}) h_{lm} \Big( \os{e}{n}^\mu_k\partial_\mu \hat{\Gamma}^m_{ij} + \cdots +\os{e}{1}^\mu_k\partial_\mu [\hat{\Gamma}^m_{ij}]_{n-1} + [\hat{\Gamma}^m_{ks}]_{n-1}  [\hat{\Gamma}^s_{ij}]_{1} +\cdots + [\hat{\Gamma}^m_{ks}]_{1}  [\hat{\Gamma}^s_{ij}]_{n-1} \Big). \quad \qquad   
 \end{eqnarray}
Such a piece takes place in the step $n$ Lagrangian (\ref{Mdollsd}) and the $n$th order expanded Lagrangian of both decompositions with the multiplication of the expanded fields of the measure. Therefore, a variation with respect to $\os{n}{n}_t$ which acts as a Lagrange multiplier results in (\ref{norderpiece}).
It consists of second order derivations of $n$th order fields or multiplication of the fields whose orders combine to give $n$. As seen from the definition of $[\hat{R}_n]$  (\ref{hatRn}), it will receive contributions which will lead to simplifications.

\section{Conclusion}
\label{discuss}

There are various approximation schemes to study GR. Among them the most successful one is the PN expansion which requires a weak field assumption together with slow velocity approximation. The $1/c$  expansion of GR is another NR approximation scheme and it differs from the PN expansion as it captures strong fields. The recent paper \cite{Elbistan:2022plu} studied the large $c$ expansion of GR within a particular $3+1$ decomposition, named the KS decomposition up to $c^{-2}$ order which corresponds to 1PN order in PN framework. In addition, equations of motion and the symmetries of the expanded Lagrangians were found.

In this paper, we have revisited the $1/c$ expansion of GR in a unified perspective. As being dual to each other, both the ADM and the KS decompositions have the same form of the Einstein-Hilbert action (\ref{SEH}). Following the idea of this duality, we first develop an expansion scheme for that parent action. Then, we perform the expansion within our novel framework up to $c^{-3}$ order systematically without specifying any decomposition. Since the ADM and the KS decompositions are dual to each other before the expansion, it is natural to ask whether this duality survives after the expansion. This is the first question that we pursue here and the answer is positive. Similar to the unexpanded Einstein-Hilbert action (\ref{SEH}), both decompositions have a common form of expansion in our step formalism. Since the KS decomposition was worked previously in \cite{Elbistan:2022plu}, we applied our formalism to the ADM decomposition and expand it up to $c^{-3}$ order, which goes beyond 1PN order as it captures terms from higher PN orders. In order to demonstrate the duality at the level of expansion, we applied our technique to the KS decomposition up to $c^{-1}$. We also make some all order observations for the expanded parent Lagrangian. We conclude that, although they are dual to each other, the ADM decomposition is at least technically better suited for $1/c$ expansion. This is especially because $\os{e}{n}_i=0$ for any $n\geq 1$ in ADM decomposition (\ref{ADMchoice}). In addition, its $K_{ij}$ tensor is symmetric whereas its counterpart in the KS decomposition contains both symmetric and anti-symmetric pieces.

As it stands, $M$ or $N$ (or collectively the field $n_t$) plays the role of Lagrange multiplier as no derivative acts on. In our Appendix, we provide a conformal redefinition of the field $M \to \psi$ in order to obtain the Poisson equation in the Newtonian limit. Other than that, we have not investigated the expanded equations of motion and the symmetries of the ADM decomposition.  As a future work, we would like to generate equations of motion for the $c^{-3}$ order ADM Lagrangian.  Our findings can be adopted to $1/c^2$ expansion which contains only even order terms. In that case, we can easily obtain $c^{-6}$ order effective ADM Lagrangian. In our Appendix \ref{Poisson} we made a brief introduction to the even expansion.

In this study, we mainly deal with developing a covariant expansion scheme rather than applying it. 
The natural field of application of our findings are strong gravitational systems like dense rotating neuton stars \cite{HartleThorn} and black hole mergers \cite{Abbott},  but it would be also interesting to work out condensed matter applications. In \cite{Cvetic:2012vg}, the authors discussed the Zermelo and Randers form of the black hole and graphene metric which correspond to ADM and  KS decompositions, respectively. In \cite{Gibbons:2008zi}, the authors considered the relation between the Zermelo and Randers metrics within the Finslerian geometry. As any spatial dimension $d$ can be adapted in our formalism, we may reserve such potential applications for a follow-up study.

\section*{Acknowledgment}

I am grateful to Erdal Çatak and Mustafa Mullahasanoğlu for their comments on the manuscript. 
\noindent M. E. is supported by İstanbul Bilgi University research fund (BAP) with grant no: 2024.01.009 and by The Scientific and Technological Research Council of Turkey (TÜBİTAK) under grant number 125F021.

\appendix

\section*{Appendices}

\section{Decomposition of the Kerr metric} 
\label{Kerrdec}

We first consider the Kerr metric just to illustrate our decompositions. In Boyer-Lindquist coordinates  $t, r, \theta, \phi $ the Kerr metric takes the form
\begin{equation}
\label{Kerrm}
ds^2 = - c^2\left( 1 - \frac{2mr}{\Sigma}  \right) dt^2 - \frac{4 mr a\sin^2\theta}{\Sigma}c dt d\phi + \frac{A\sin^2\theta}{\Sigma} d\phi^2 + \frac{\Sigma}{\Delta_r} dr^2 +\Sigma d\theta^2 , 
\end{equation}
where 
\begin{equation}
\Sigma = r^2 + a^2 \cos^2\theta, \quad  \Delta_r = r^2 - 2mr + a^2, \quad  A = (r^2 + a+2)^2 -\Delta_r a^2 \sin^2\theta .
\end{equation}
Above, $m$ is the mass and $a$ is the ratio between the angular momentum and $m$. 
Note that none of the metric components depend on the coordinate $t$. 

ADM coordinates (\ref{ADMm}) leads to the following identification of the fields
\begin{equation}
\label{ADMKerrf}
N= \sqrt{1- \frac{2mr}{\Sigma} + \frac{4m^2 r^2 a^2}{\Sigma A}}, \quad  N^\phi = -\frac{2mra}{A}, \quad h_{rr} = \frac{\Sigma}{\Delta_r}, \quad h_{\theta\theta} = \Sigma, \quad h_{\phi\phi} = \frac{A\sin^2\theta}{\Sigma}, 
\end{equation}
where all other components like $N^\theta$, $h_{r\phi}$ etc. are vanishing. The zeroth step Lagrangian $[L]_0 = [R]_0 [\sqrt{-g}]_0$ (\ref{Mdollsa}) in our Matryosha expansion consists of 
\begin{eqnarray}
  [\sqrt{-g}]_0 &=& N \sqrt{h} =  \sqrt{1- \frac{2Mr}{\Sigma} + \frac{4M^2 r^2 a^2}{\Sigma A}} \sqrt{\frac{A\sin^2\theta \Sigma }{\Delta_r}}, 
\end{eqnarray} 
and $[R]_0$ (\ref{step0L}) which contains spatial Christoffel symbols $\Gamma^{i}_{jk}$ and their spatial covariant derivatives $D_k \Gamma^n_{ij}$ as well as the tensor $K_{ij}$. 

On the other hand, the KS decomposition (\ref{KSm}) leads to following non-vanishing fields,
\begin{equation}
\label{KSKerrf}
M = \sqrt{1 - \frac{2mr}{\Sigma}},  \quad M_\phi=  -\frac{2 m r a \sin^2\theta}{\Sigma - 2mr}, \quad h_{rr} = \frac{\Sigma}{\Delta_r}, \quad  h_{\theta\theta}= \Sigma, \quad  h_{\phi\phi}= \frac{4m^2r^2 a^2 \sin^2\theta + A\sin^2\theta(\Sigma- 2mr)}{\Sigma(\Sigma- 2 mr)}.
\end{equation}

We refer to \cite{Ergen:2020yop} for a more detailed discussion.

\section{Comment on the time derivatives in our steps}
\label{tdersteps}

The Kerr metric (\ref{Kerrm}) admits a Killing vector $\partial_t$ and the related zeroth step Lagrangian (\ref{step0L}) contains no time derivatives. This is independent from the fact that ADM (\ref{ADM0}) or KS zeroth order Lagrangians are free of time derivatives. In $1/c$ expansion the zeroth order Lagrangians are always stationary and they source the time dependent fields in higher orders. 

But, in principle, our step zero Lagrangian within Matryosha expansion (\ref{step0L}) includes two time derivatives. This can be seen most easily in the ADM decomposition. Let us consider (\ref{step0L}) in ADM coordinates (\ref{ADMchoice}). Ricci tensor $R$ does not have any time derivative in this particular choice. Derivatives appear only in the second part 
\begin{eqnarray}
- (h^{ij}h^{kl} - h^{ik}h^{jl}) K_{ij} K_{kl} &=& -N^{-2}(h^{ij}h^{kl} - h^{ik}h^{jl}) (\frac{1}{2}\dot{h}_{ij} - D_{(i} N_{j)}) (\frac{1}{2}\dot{h}_{kl} - D_{(k} N_{l)})  \\
&=& -\frac{N^{-2}}{4} ( h^{ij}\dot{h}_{ij}h^{kl}\dot{h}_{kl} - \dot{h}_{ij}\dot{h}_{kl} h^{ik}h^{jl} ) \nonumber \\
&+&N^{-2} (h^{ij}\dot{h}_{ij} D_k N^{k} - \dot{h}_{ij} D^{i} N^{j}) \nonumber\\
&-& N^{-2} \big( (D_i N^i)^2 - D_{(i} N_{j)} D^{(i} N^{j)}   \big). \nonumber 
\end{eqnarray}      
As it is seen easily, the last part does not contain any derivatives. The middle term contains only one time derivative. And, the first term containing only spatial metric and its derivatives has two time derivatives. 

This is also true for the KS coordinates. Therefore, our zeroth step Lagrangian (\ref{step0L}) has terms containing zero, one and two time derivatives in general and it contributes to three different orders. The expansion procedure (\ref{KSexp}) or (\ref{ADMexp}) adds no time derivatives to the system, but includes expanded fields. Those time derivatives are always contained in the further steps of the expansion e.g., (\ref{step1Lfin}). So, at each step we have zero, one and two time derivative terms and the related expanded fields. Thus, at zeroth step we have $c^{0}, c^{-1}, c^{-2}$ terms, at first step we have $c^{-1}, c^{-2}, c^{-3}$ etc.

\section{Conformal redefinition, even expansion and the Poisson equation}
\label{Poisson}

In this section, we would like to obtain the NR Poisson equation in the Newtonian limit. Recalling that $n_t$ acts as a Lagrange multiplier, we first make the following redefinition
\begin{equation}
n_t = e^{\psi/2},  \quad  h_{ij} = e^{-\psi}\gamma_{ij},
\end{equation}
where the new field $\psi(t, \vec{x})$ replaces the scalar $n_{t}$ and $\gamma_{ij}(t, \vec{x})$ is the new spatial metric. When applied to ADM coordinates, the metric decomposition becomes
\begin{equation}
ds^2 = -c^2 e^\psi dt^2 + e^{-\psi}\gamma_{ij} (dx^i + c N^i dt)^2,
\end{equation}
cf. (\ref{ADMm}) where $M=e^{\psi/2}$.  On the other hand, the KS metric is modified as
\begin{equation}
ds^2 = -e^\psi (cdt + M_i dx^i)^2 + e^{-\psi}\gamma_{ij} dx^i dx^j,
\end{equation}
with the substitution $N= e^{\psi/2}$. 

As we will show, in these new coordinates the relation to Newtonian limit and Poisson equation is explicit. The new scalar field $\psi(t, \vec{x})$ plays the role of the Newtonian potential. We will also prove that our step/Matryoshka framework can be adapted to new coordinates without any problem. Firstly, we show that our identity (\ref{covid}) is stil valid. The Christoffel symbols $\hat{\Gamma}^i_{jk}(h)$ (\ref{Christoffel}) are defined as
\begin{equation}
\label{Christoffelc}
\hat{\Gamma}^i_{jk}(h) = \hat{\Gamma}^i_{jk}(\gamma) - \frac{1}{2} \gamma^{mk} (\gamma_{kj}\hat{\partial}_i\psi  + \gamma_{ik}\hat{\partial}_j\psi - \gamma_{ij}\hat{\partial}_k\psi ),
\end{equation}  
where $\gamma^{ij}(t,\vec{x})$ is the inverse metric i.e.,  $\gamma^{ik}\gamma_{kj}=\delta^i_j$ and 
\begin{equation}
\hat{\Gamma}^k_{ij}(\gamma) = \frac{1}{2} \gamma^{km} (\hat{\partial}_i \gamma_{mj} + \hat{\partial_j } \gamma_{mi} - \hat{\partial}_m \gamma_{ij}),
\end{equation}
is the new Christoffel symbol. A straightforward computation shows that the identity (\ref{covid}) implies
\begin{equation}
\hat{D}_i(\gamma) \gamma_{jk} \equiv \hat{\partial}_i \gamma_{jk} - \hat{\Gamma}^m_{ij}(\gamma)\gamma_{mk} - \hat{\Gamma}^m_{ik}(\gamma)\gamma_{mj} =0. 
\end{equation}
Therefore, the expansion identities in our Section \ref{identities} works well in those conformal coordinates. On the other hand, the extra piece in (\ref{Christoffelc}) will induce new terms in the Ricci scalar $\hat{R}$ (\ref{hatR0}) as
\begin{equation}
\hat{R} = h^{ij}\hat{R}_{ij} = \gamma^{ij}\Big( \hat{R}_{ij} (\gamma) + 2\hat{D}_i \hat{D}_j \psi - \frac{1}{2}\hat{D}_i\psi \hat{D}_j \psi \Big), 
\end{equation}
where $\hat{R}_{ij} (\gamma) $ is nothing but (\ref{hatR0}) with $h_{ij} \to \gamma_{ij}$. The related piece of the Einstein-Hilbert action (\ref{SEH}) is generic for both decompositions
\begin{equation}
n_t \sqrt{h} h^{ij}\hat{R}_{ij} = \sqrt{\gamma} \gamma^{ij} (\hat{R}_{ij} (\gamma) + 2\hat{D}_i \hat{D}_j \psi - \frac{1}{2}\hat{D}_i\psi \hat{D}_j \psi). 
\end{equation}

For the sake of simplicity let us focus on the ADM decomposition. Then, the second expression of the integrand with $2\hat{D}_i \hat{D}_j \psi$ turns out to be a boundary term and vanishes under integral.

The modifications in the $K_{ij}$ tensor and its contraction depends heavily on the decomposition. In ADM coordinates, the $K_{ij}$ (\ref{KADM}) turns out to be 
\begin{equation}
K_{ij} = e^{-\psi/2} \left(   \frac{e^{-\psi}}{2} (\dot{\gamma}_{ij} - \gamma_{ij}) - D_{(i}N_{j)} - N_{(i} \hat{\partial}_j \psi) - \frac{1}{2} \gamma_{ij} N^k\hat{\partial}_k\psi\right).  
\end{equation}

The direct route to NR limit passes through the $\frac{1}{c^2}$ (even) expansion (see e.g., \cite{Elbistan:2022plu}) which is a consistent truncation of our $\frac{1}{c}$ expansion scheme (\ref{ADMexp}). The even expansion procedure contains the even order powers of scalar $\psi$ and metric $\gamma_{ij}$ but odd powers of $N^i$ such that
\begin{equation}
\psi = \psi + \frac{1}{c^2}\os{\psi}{2} + \cdots , \quad  N^i = \frac{1}{c} \os{N}{1}^{i} + \frac{1}{c^3}\os{N}{3}^{i}+\cdots , \quad \gamma_{ij} = \gamma_{ij} + \frac{1}{c^2} \os{\gamma}{2}_{ij}+\cdots .
\end{equation}
This expansion generates an even order expansion of the Lagrangian such that 
\begin{equation}
L_{ADM} = \os{L}{0}_{ADM} + \frac{1}{c^2}\os{L}{2}_{ADM}+\cdots ,
\end{equation}
and $\os{L}{n}_{ADM}$ can be computed through our procedure presented in Sections \ref{identities}, \ref{Ricciexpansion} and \ref{stepsexp}. Going over the same steps, we compute the zeroth order ADM Lagrangian $ \os{L}{0}_{ADM}$ as
\begin{equation}
 \os{L}{0}_{ADM}^{even} = \sqrt{\gamma} \gamma^{ij}\Big(R_{ij} + 2 D_i D_j \psi - \frac{1}{2}(D_i\psi)(D_j \psi)\Big).  
\end{equation}
Pay attention to the fact that in even expansion $ \os{L}{0}_{ADM}^{even}$ does not receive contributions from $N^i$ terms as they belong to higher orders, cf. (\ref{L0ADM}). The Ricci tensor $R_{ij}(\gamma)$ depends solely on $\gamma_{ij}$ and $ \sqrt{\gamma} \gamma^{ij}D_i D_j \psi$  is just a boundary term under integration. Therefore, the variation w.r.t. $\psi$ field yields
\begin{equation}
\frac{\delta}{\delta \psi}  \os{L}{0}_{ADM}^{even} = D^i D_i \psi , 
\end{equation}
and the variation of the matter Lagrangian is defined to be the density i.e., $\frac{\delta}{\delta \psi} L_{m} \equiv \rho$.  Together, they directly lead to the Poisson equation. Therefore, the field $\psi$ in ADM decomposition has a direct physical meaning: it is the analogue of Newtonian potential. 

\section{Relation to Newton-Cartan theory} 
\label{NC}

Like PN expansion, the $3+1$ formulation of the large $c$ expansion \cite{Elbistan:2022plu} is based on the Newton-Cartan theory. Consider a normalized, time-like vector field $u^\mu$ such that $g_{\mu\nu} u^\mu u^\nu =-1$ and $e_i =e_i^\mu\partial_\mu $ being orthogonal to it i.e., $g_{\mu\nu} u^\mu e_i^\nu =0$. Out of that, we can form a dual frame $(n, e^i)$ satisfying $e^i_\mu e^\mu_j =\delta^i_j$, $u^\mu e^i_\mu =0$ and $e^\mu_i e^j_\mu= \delta^j_i$. This structure allows us to define the identity as
\begin{equation}
-u^\mu n_\nu + e^\mu_i e^i_\nu = \delta^\mu_\nu. 
\end{equation}  
Thus, the Newton-Cartan split of a vector $A_\mu$ is
\begin{equation}
\label{split1}
A_\mu = -(A_\nu u^\nu)n_\mu + (A_\nu e^\nu_i)e^i_\mu,
\end{equation}
and the Newton-Cartan split of an arbitrary tensor $B_{\mu\nu}$ is 
\begin{equation}
\label{split2}
B_{\mu\nu} = (B_{\alpha\beta})n_\mu n_\nu - (B_{\alpha\beta}u^\alpha e^\beta_j)n_\mu e^j_\nu - (B_{\alpha\beta}e^\alpha_i u^\beta)e^i_\mu n_\nu + (R_\alpha\beta e^\alpha_i e^\beta_j) e^i_\mu e^j_\nu .   
\end{equation}
Lastly, we define the spatial metric and its inverse in terms of frame vectors as $h_{ij} = g_{\mu\nu} e_i^\mu e^\nu_j$ and its inverse as $h^{ij} = g^{\mu\nu}e^i_\mu e^j_\nu$. 

Now, for the well known ADM decomposition (\ref{ADMm}), one can easily find that 
\begin{equation}
e^\mu_i \partial_\mu = \partial_i,  \quad  n_\mu dx^\mu = -N dt, \quad  e^i_\mu dx^\mu = dx^i + N^i dt, \quad u^\mu\partial_\mu = N^{-1}(\partial_t - N^i\partial_i). 
\end{equation}

Similar computations can be performed for the KS decomposition (\ref{KSm}) and one can observe that those two are dual to each other. ADM or KS, the Ricci tensor and the Ricci scalar takes the common form (\ref{SEH}) once the Newton-Cartan split (\ref{split1}) and (\ref{split2}) are applied.  

Therefore, our fields $N(t, \vec{x})$ etc. can be identified via Newton-Cartan framework. We refer to \cite{Elbistan:2022plu} for more details.

\end{document}